\documentstyle[preprint,tighten,prc,aps,epsf]{revtex}

\begin{document}
\draft

\def\simge{\hspace*{0.2em}\raisebox{0.5ex}{$>$}
     \hspace{-0.8em}\raisebox{-0.3em}{$\sim$}\hspace*{0.2em}}
\def\simle{\hspace*{0.2em}\raisebox{0.5ex}{$<$}
     \hspace{-0.8em}\raisebox{-0.3em}{$\sim$}\hspace*{0.2em}}
\def\bra#1{{\langle#1\vert}}
\def\ket#1{{\vert#1\rangle}}
\def\coeff#1#2{{\scriptstyle{#1\over #2}}}
\def\undertext#1{{$\underline{\hbox{#1}}$}}
\def\hcal#1{{\hbox{\cal #1}}}
\def\sst#1{{\scriptscriptstyle #1}}
\def\eexp#1{{\hbox{e}^{#1}}}
\def\rbra#1{{\langle #1 \vert\!\vert}}
\def\rket#1{{\vert\!\vert #1\rangle}}
\def\lsim{{ <\atop\sim}}
\def\gsim{{ >\atop\sim}}
\def\nubar{{\bar\nu}}
\def\psibar{{\bar\psi}}
\def\Gmu{{G_\mu}}
\def\alr{{A_\sst{LR}}}
\def\wpv{{W^\sst{PV}}}
\def\evec{{\vec e}}
\def\notq{{\not\! q}}
\def\notl{{\not\! \ell}}
\def\notk{{\not\! k}}
\def\notp{{\not\! p}}
\def\notpp{{\not\! p'}}
\def\notder{{\not\! \partial}}
\def\notcder{{\not\!\! D}}
\def\notA{{\not\!\! A}}
\def\notv{{\not\!\! v}}
\def\Jem{{J_\mu^{em}}}
\def\Jana{{J_{\mu 5}^{anapole}}}
\def\nue{{\nu_e}}
\def\mn{{m_\sst{N}}}
\def\mns{{m^2_\sst{N}}}
\def\me{{m_e}}
\def\mes{{m^2_e}}
\def\mq{{m_q}}
\def\mqs{{m_q^2}}
\def\mz{{M_\sst{Z}}}
\def\mzs{{M^2_\sst{Z}}}
\def\ubar{{\bar u}}
\def\dbar{{\bar d}}
\def\sbar{{\bar s}}
\def\qbar{{\bar q}}
\def\sstw{{\sin^2\theta_\sst{W}}}
\def\gv{{g_\sst{V}}}
\def\ga{{g_\sst{A}}}
\def\pv{{\vec p}}
\def\pvs{{{\vec p}^{\>2}}}
\def\ppv{{{\vec p}^{\>\prime}}}
\def\ppvs{{{\vec p}^{\>\prime\>2}}}
\def\qv{{\vec q}}
\def\qvs{{{\vec q}^{\>2}}}
\def\xv{{\vec x}}
\def\xpv{{{\vec x}^{\>\prime}}}
\def\yv{{\vec y}}
\def\tauv{{\vec\tau}}
\def\sigv{{\vec\sigma}}
\def\sst#1{{\scriptscriptstyle #1}}
\def\gpnn{{g_{\sst{NN}\pi}}}
\def\grnn{{g_{\sst{NN}\rho}}}
\def\gnnm{{g_\sst{NNM}}}
\def\hnnm{{h_\sst{NNM}}}

\def\xivz{{\xi_\sst{V}^{(0)}}}
\def\xivt{{\xi_\sst{V}^{(3)}}}
\def\xive{{\xi_\sst{V}^{(8)}}}
\def\xiaz{{\xi_\sst{A}^{(0)}}}
\def\xiat{{\xi_\sst{A}^{(3)}}}
\def\xiae{{\xi_\sst{A}^{(8)}}}
\def\xivtez{{\xi_\sst{V}^{T=0}}}
\def\xivteo{{\xi_\sst{V}^{T=1}}}
\def\xiatez{{\xi_\sst{A}^{T=0}}}
\def\xiateo{{\xi_\sst{A}^{T=1}}}
\def\xiva{{\xi_\sst{V,A}}}

\def\rvz{{R_\sst{V}^{(0)}}}
\def\rvt{{R_\sst{V}^{(3)}}}
\def\rve{{R_\sst{V}^{(8)}}}
\def\raz{{R_\sst{A}^{(0)}}}
\def\rat{{R_\sst{A}^{(3)}}}
\def\rae{{R_\sst{A}^{(8)}}}
\def\rvtez{{R_\sst{V}^{T=0}}}
\def\rvteo{{R_\sst{V}^{T=1}}}
\def\ratez{{R_\sst{A}^{T=0}}}
\def\rateo{{R_\sst{A}^{T=1}}}

\def\mro{{m_\rho}}
\def\mks{{m_\sst{K}^2}}
\def\mpi{{m_\pi}}
\def\mpis{{m_\pi^2}}
\def\mom{{m_\omega}}
\def\mphi{{m_\phi}}
\def\Qhat{{\hat Q}}

\def\FOS{{F_1^{(s)}}}
\def\FTS{{F_2^{(s)}}}
\def\GAS{{G_\sst{A}^{(s)}}}
\def\GES{{G_\sst{E}^{(s)}}}
\def\GMS{{G_\sst{M}^{(s)}}}
\def\GATEZ{{G_\sst{A}^{\sst{T}=0}}}
\def\GATEO{{G_\sst{A}^{\sst{T}=1}}}
\def\mdax{{M_\sst{A}}}
\def\mustr{{\mu_s}}
\def\rsstr{{r^2_s}}
\def\rhostr{{\rho_s}}
\def\GEG{{G_\sst{E}^\gamma}}
\def\GEZ{{G_\sst{E}^\sst{Z}}}
\def\GMG{{G_\sst{M}^\gamma}}
\def\GMZ{{G_\sst{M}^\sst{Z}}}
\def\GEn{{G_\sst{E}^n}}
\def\GEp{{G_\sst{E}^p}}
\def\GMn{{G_\sst{M}^n}}
\def\GMp{{G_\sst{M}^p}}
\def\GAp{{G_\sst{A}^p}}
\def\GAn{{G_\sst{A}^n}}
\def\GA{{G_\sst{A}}}
\def\GETEZ{{G_\sst{E}^{\sst{T}=0}}}
\def\GETEO{{G_\sst{E}^{\sst{T}=1}}}
\def\GMTEZ{{G_\sst{M}^{\sst{T}=0}}}
\def\GMTEO{{G_\sst{M}^{\sst{T}=1}}}
\def\lamd{{\lambda_\sst{D}^\sst{V}}}
\def\lamn{{\lambda_n}}
\def\lams{{\lambda_\sst{E}^{(s)}}}
\def\bvz{{\beta_\sst{V}^0}}
\def\bvo{{\beta_\sst{V}^1}}
\def\Gdip{{G_\sst{D}^\sst{V}}}
\def\GdipA{{G_\sst{D}^\sst{A}}}
\def\fks{{F_\sst{K}^{(s)}}}
\def\FIS{{F_i^{(s)}}}
\def\fpi{{F_\pi}}
\def\fk{{F_\sst{K}}}

\def\RAp{{R_\sst{A}^p}}
\def\RAn{{R_\sst{A}^n}}
\def\RVp{{R_\sst{V}^p}}
\def\RVn{{R_\sst{V}^n}}
\def\rva{{R_\sst{V,A}}}
\def\xbb{{x_B}}

\def\mlq{{M_\sst{LQ}}}
\def\mlqs{{M_\sst{LQ}^2}}
\def\lscal{{\lambda_\sst{S}}}
\def\lvect{{\lambda_\sst{V}}}

\def\PR#1{{{\em   Phys. Rev.} {\bf #1} }}
\def\PRC#1{{{\em   Phys. Rev.} {\bf C#1} }}
\def\PRD#1{{{\em   Phys. Rev.} {\bf D#1} }}
\def\PRL#1{{{\em   Phys. Rev. Lett.} {\bf #1} }}
\def\NPA#1{{{\em   Nucl. Phys.} {\bf A#1} }}
\def\NPB#1{{{\em   Nucl. Phys.} {\bf B#1} }}
\def\AoP#1{{{\em   Ann. of Phys.} {\bf #1} }}
\def\PRp#1{{{\em   Phys. Reports} {\bf #1} }}
\def\PLB#1{{{\em   Phys. Lett.} {\bf B#1} }}
\def\ZPA#1{{{\em   Z. f\"ur Phys.} {\bf A#1} }}
\def\ZPC#1{{{\em   Z. f\"ur Phys.} {\bf C#1} }}
\def\etal{{{\em   et al.}}}

\def\delalr{{{delta\alr\over\alr}}}
\def\pbar{{\bar{p}}}
\def\lamchi{{\Lambda_\chi}}

\def\qw0{{Q_\sst{W}^0}}
\def\qwp{{Q_\sst{W}^P}}
\def\qwn{{Q_\sst{W}^N}}
\def\qwe{{Q_\sst{W}^e}}

\def\gae{{g_\sst{A}^e}}
\def\gve{{g_\sst{V}^e}}
\def\gvf{{g_\sst{V}^f}}
\def\gaf{{g_\sst{A}^f}}
\def\gvu{{g_\sst{V}^u}}
\def\gau{{g_\sst{A}^u}}
\def\gvd{{g_\sst{V}^d}}
\def\gad{{g_\sst{A}^d}}

\def\gvftil{{\tilde g_\sst{V}^f}}
\def\gaftil{{\tilde g_\sst{A}^f}}
\def\gvetil{{\tilde g_\sst{V}^e}}
\def\gaetil{{\tilde g_\sst{A}^e}}
\def\gvqtil{{\tilde g_\sst{V}^e}}
\def\gaqtil{{\tilde g_\sst{A}^e}}
\def\gvutil{{\tilde g_\sst{V}^e}}
\def\gautil{{\tilde g_\sst{A}^e}}
\def\gvdtil{{\tilde g_\sst{V}^e}}
\def\gadtil{{\tilde g_\sst{A}^e}}

\def\hvf{{h_\sst{V}^f}}
\def\hvu{{h_\sst{V}^u}}
\def\hvd{{h_\sst{V}^d}}
\def\hve{{h_\sst{V}^e}}
\def\hvq{{h_\sst{V}^q}}

\def\delp{{\delta_P}}
\def\delzp{{\delta_{00}}}
\def\deld{{\delta_\Delta}}
\def\dele{{\delta_e}}

\def\apv{{A_\sst{PV}}}
\def\apvnsid{{A_\sst{PV}^\sst{NSID}}}
\def\qpv{{Q_\sst{W}}}

\title{
\hfill{\normalsize INT \#DOE/ER/40561-37-INT98}\\[2.5cm]
Low-Energy Parity-Violation and New Physics}

\author{M.J. Ramsey-Musolf$^{a,b,c}$
\\[0.3cm]
}
\address{
$^a$ Department of Physics, University of Connecticut,
Storrs, CT 06269 USA\\
$^b$ Institute for Nuclear Theory,
University of Washington, Seattle, WA 98195 USA\\
$^c$ Theory Group, Thomas Jefferson National Laboratory, Newport News,
VA 23606 USA
}


\maketitle

\begin{abstract}
The new physics sensitivity of a variety of low-energy parity-violating (PV)
observables is analyzed. A comparison is made between atomic
PV for a single isotope, atomic PV using isotope ratios, and PV electron-hadron
and electron-electron scattering. The complementarity among these observables,
as well as with high-energy processes, is emphasized. Theoretical uncertainties
entering the interpretation of low-energy measurements are discussed.

\end{abstract}


\narrowtext

\section{Introduction}
\label{sec:intro}

Low-energy parity-violating (PV) observables have played an important role
in uncovering the structure of the electroweak sector of the Standard Model.
Now that the predictions of the Standard Model have been tested and
confirmed at
the one-loop level over a wide range of processes and energies \cite{Sch95},
attention has turned to the search for physics beyond the
Standard Model. In this regard, low-energy parity-violation continues to
provide important information. As has been noted by several
authors\cite{Ros98,New98,Nel97},
the recent precise determination of the cesium weak charge, $\qpv$, in an
atomic parity-violation
(APV) experiment performed by
the  Boulder group\cite{Woo97} places stringent constraints on a variety of
new physics
scenarios. The importance of this benchmark measurement is reflected, in
part, by
the efforts of other experimental groups to determine $\qpv$ for cesium as
well as
other atoms\cite{Bou98,For93,Bud98}. Future improvements in the APV
sensitivity to
new physics poses a challenge to both atomic experimentalists and theorists.
Indeed, given the experimental precision reported by the Boulder group,
atomic theory error now
constitutes the dominant uncertainty associated with the intepretation of
atomic PV (APV) observables. Whether this atomic theory error can be
reduced to the level
of the experimental uncertainty remains to be seen. An experimental
strategy for circumventing
the atomic theory uncertainty is to measure PV observables for different
atoms along an
isotope chain. Standard Model predictions for {\em ratios} of such
observables are largely
atomic theory-independent. Consequently, several groups have undertaken APV
isotope ratio
measurements in the hopes of minimizing the impact of atomic theory
uncertainties on the
extraction of new physics constraints\cite{For93,Bud98,Dem95}.

Historically, the use of polarized electrons produced in accelerator
experiments has,
along with APV, played a part in testing the Standard
Model\cite{Pre78,Hei89,Sou90}. In the
past decade, however, PV electron scattering (PVES) has received less attention
than APV in this respect since (a) the experimental precision achievable
with APV has improved markedly and (b) interest in PVES has focused on its
use in probing the nucleon's $s\bar{s}$ sea. The interest in nucleon
strangeness has spawned
a program of experiments at MIT-Bates, Mainz, and the Jefferson
Laboratory to measure the left-right asymmetry, $\alr$, on a variety of
targets\cite{PVES89,Mue97,Ani98}.  Recently, the attention of the PVES
community has returned
to the use of these experiments to probe new physics\cite{PV97}. In the purely
leptonic sector, work on a high-precision PV M\" oller scattering experiment
has begun at SLAC\cite{SLAC97}. In addition, a program of \lq\lq second
generation" PVES
experiments -- designed to look for physics beyond the Standard Model -- is
under consideration for the Jefferson Lab. The feasibility of such PVES new
physics searches
stems, in part, from the high luminosity and remarkably stable and clean
electron beam
achieved by the CEBAF accelerator\cite{Sou98}.

Although there have appeared numerous discussions of $\qpv($cesium) in the
literature recently, relatively little attention has been paid to the other
low-energy PV observables mentioned above. In this paper, we therefore
consider the new physics sensitivities of APV isotope ratios and PVES
asymmetries, making comparison with sensitivities of $\qpv$ and high-energy
observables. In doing so,  we focus
on \lq\lq direct" new physics, that is, extensions of the Standard Model
which manifest themselves at low-energies as new four-fermion contact
interactions. The sensitivity of APV and PVES to \lq\lq oblique" new physics
has been discussed
elsewhere\cite{Ros98,Pes90,Mar90,Mus94,Muk98}\footnote{Previously,
isotope ratios were shown to display a significantly different sensitivity
to oblique new physics
than does $\qpv$ for a single isotope. We show that a similar situation
holds in the case of
direct new physics.}. After quantifying the generic
new physics sensitivities of PV observables, we specify to a variety of
models in
order to illustrate the complementarity of prospective measurements. In
particular, we show that
to leading order, the elastic $ep$ asymmetry, $\alr(^1{\hbox{H}})$ ,and APV
isotope ratios, ${\cal
R}$, are sensitive to the same combination of possible new interactions.
These two observables,
while subject to different systematic and theoretical corrections, provide
the same window on
direct new physics. We also find that a 2-3\% determination of
$\alr(^1{\hbox{H}})$  would improve
the new physics reach of low-energy PV by nearly a factor of two over the
present
cesium APV sensitivity. A similar improvement would obtain if the present
cesium atomic theory
error were improved by a factor of four. Apart from the M\" oller
asymmetry, the remaining
asymmetries display a smaller new physics reach than $\qpv$,
$\alr(^1{\hbox{H}})$ , or ${\cal
R}$.   In illustrating model variations on the general
pattern of new physics sensitivity, we consider additional neutral gauge
bosons, leptoquarks, and
fermion compositeness. We  also discuss the sensitivity of PV observables
to R-parity violating
supersymmetric interactions and compare this sensitivity with up-dated
bounds from
super-allowed $\beta$-decay.

The use of low-energy measurements to probe new physics requires that
conventional, many-body
physics associated with atoms and nuclei be sufficiently well-understood.
From this standpoint,
we show that, in principle, PVES provides the theoretically \lq\lq
cleanest" new physics
probe. This feature is most apparent for PV M\" oller scattering, as it is
a purely leptonic
process. In the case of semi-leptonic PV observables, the reason for
minimal theoretical
uncertainty is two-fold: (a)
$\alr$ depends on a ratio of electroweak amplitudes, from which the largest
hadronic
effects cancel, leaving essentially a dependence on $\qpv$ of the target
nucleus; and (b) the
largest remaining hadronic corrections to this cancellation can be
separated from
$\qpv$ and measured by exploiting their kinematic dependence. Consequently,
the dominant
uncertainty in the interpretation of PVES new physics studies is likely to
be experimental.
We illustrate these features in the case of $\alr(^1{\hbox{H}})$ and
discuss the kinematics
to make such a clean separation of $\qpv$ feasible.

The situation in the case of APV differs from that of PVES.
The atomic theory uncertainty associated with extracting $\qpv$ from cesium
APV is
about four times larger than the experimental error. This situation has
prompted the
consideration of the isotope ratios
${\cal R}$, from which the dominant atomic theory uncertainties cancel.
Unfortunately,
${\cal R}$ carries a problematic sensitivity to changes in the neutron
distribution,
$\rho_n(r)$, from one nucleus to the next along an isotope change.
Following on the earlier work of Refs. \cite{For90,Pol92,Che93}, we analyze
the impact which
the uncertainties in the neutron distribution,
$\rho_n(r)$, have on the extraction of new physics limits from ${\cal R}$.
We consider several
atoms presently under experimental consideration and quantify the level of
$\rho_n$ uncertainty
acceptable in order for relevant new physics limits to be obtained from
${\cal R}$.

Our discussion of these issues is organized as follows. In Section II, we
outline our
conventions and definitions, and in Section III discuss general new physics
sensitivities
of low-energy PV observables. In Section IV we illustrate these
sensitivities for different
new physics scenarios. Section V contains an analysis of theoretical
uncertainties. A discussion
of kinematic considerations for a prospective PV elastic $ep$ experiment is
also included. In
Section VI we summarize our conclusions.

\section{New Physics and the Weak Charge}
\label{sec:formal}

For each PV observable, the quantity of interest here is the weak charge
$\qpv$ of the nucleus (electron), which characterizes the strength of the
electron axial vector $\times$ nucleus (electron) vector weak neutral current
interaction:
\begin{equation}
\qpv=\qw0+\Delta\qpv \ \ \ .
\end{equation}
Here, $\qw0$ gives the contribution in the Standard Model while
$\Delta\qpv$ indicates possible contributions from new interactions.
We consider $\qpv$ to be generated by the low-energy effective
Lagrangian
\begin{equation}
{\cal L}={\cal L}^\sst{PV}_\sst{S.M.}+{\cal L}^\sst{PV}_\sst{NEW}\ \ \ ,
\end{equation}
where
\begin{eqnarray}
\label{eq:lsmodel}
{\cal L}^\sst{PV}_\sst{S.M.}&=&{G_F\over 2\sqrt{2}}\gae{\bar e}\gamma_\mu
	\gamma_5 e\sum_f \gvf{\bar f}\gamma^\mu f \\
\label{eq:lnew}
{\cal L}^\sst{PV}_\sst{NEW}&=&{4\pi\kappa^2\over\Lambda^2} {\bar e}
	\gamma_\mu\gamma_5 e\sum_f\hvf{\bar f}
	\gamma^\mu f\ \ \ .
\end{eqnarray}
Here $\gvf=2T_3^f-4Q_f\sstw$ and $\gaf=-2T_3^f$ are the tree level
Standard Model fermion-$Z^0$ couplings; $\hvf$ characterizes the
interaction of the electron axial vector current with the vector
current of fermion $f$ for a given extension of the Standard Model;
$\Lambda$ is the mass scale associated with the new physics; and
$\kappa$ sets the coupling strength. Generally speaking, strongly interacting
theories take $\kappa^2\sim 1$ while for weakly interacting extensions of the
Standard Model one has $\kappa^2\sim\alpha$. For scenarios in which
the interaction of Eq. (\ref{eq:lnew}) is generated by the exchange of a new
heavy particle between the electron and fermion, the constant
$\hvf=\gaetil\gvftil$, where $\gaetil$ ($\gvftil$) are the heavy
particle axial vector (vector) coupling to the electron (fermion).

For simplicity, we do not consider contributions to $\Delta\qpv$ arising
from new scalar-pseudoscalar or tensor-pseudotensor interactions.
We also do not consider $V(e)\times A(f)$ interactions, as they do not
contribute to $\qpv$. Although the Standard Model $V(e)\times A(f)$
interaction is suppressed due to the small value of $\gae=-1+4\sstw$,
resulting in an enhanced sensitivity to new physics of this type, one
is at present not able to extract the $V(e)\times A(f)$ amplitudes
from PV observables
with the level of experimental precision attainable for $\qpv$. Moreover,
the hadronic
axial vector current is not protected by current conservation from hadronic
effects which may cloud the interpretation of the hadronic axial vector
amplitude in terms of new physics \cite{Mus90}.

It is straightforward to write down the corrections to the weak charge of
a given system arising from ${\cal L}^\sst{PV}_\sst{NEW}$. Specifically,
we consider the nucleon and electron:
\begin{eqnarray}
\label{eq:deltaqwp}
\Delta\qwp&=&\zeta(2\hvu+\hvd)\\
\Delta\qwn&=&\zeta(\hvu+2\hvd)\\
\Delta\qwe&=&\zeta\hve\ \ \ ,
\end{eqnarray}
where
\begin{equation}
\zeta = {8\sqrt{2}\pi\kappa^2\over\Lambda^2 G_F}\ \ \ .
\end{equation}

To the extent that the couplings $\gvf$ and $\hvf$ entering $\qpv$ and
$\Delta\qpv$ are of the same order of magnitude, the fractional correction
induced by new physics is
\begin{equation}
{\Delta\qpv\over\qw0}={8\sqrt{2}\pi\over\Lambda^2 G_F}\ \ \ .
\end{equation}
A one percent determination of $\qpv$ then affords a lower bound on
the mass scale associated with new physics of
\begin{equation}
\Lambda\geq\left[{8\sqrt{2}\pi\kappa^2\over 0.01 G_F}\right]^{1/2}
\approx 20\kappa \ \ {\hbox{TeV}}\ \ \ .
\end{equation}
In short, determinations of $\qpv$ at the one percent or better level
probe new physics at the TeV scale for weakly interacting theories and
the ten TeV scale for new strong interactions.

\section{Observables}
\label{sec:observe}

In this section, we discuss some of the general features of the low-energy
PV observables used to determine $\qpv$. In particular, we consider a
general atomic PV observable for a single isotope, $\apv(N)$; ratios
involving $\apv$ for different isotopes, ${\cal R}$; and the left-right
asymmetry for scattering polarized electrons from a given target,
$\alr$. Of these, the simplest is the atomic PV observable for a single
isotope, $\apv(N)$. The nuclear spin-independent (NSID) part of this observable
is given by
\begin{equation}
\apvnsid(N)=\xi\qpv = \xi\left[\qw0+Z\Delta\qwp+N\Delta\qwn\right]\ \ \ ,
\end{equation}
where $\xi$ is an atomic structure-dependent coefficient and where
\begin{equation}
\qw0 = Z(1-4\sstw)-N
\end{equation}
at tree level and
\begin{equation}
Z\Delta\qwp+N\Delta\qwn = \zeta\left[(2Z+N)\hvu
	+(2N+Z)\hvd\right]\ \ \ .
\end{equation}
A determination $\xi$
generally requires theoretical knowledge of the relevant atomic
wavefunction and, therefore, introduces theoretical uncertainty into the
extraction of $\qpv$. The relative sensitivity of $\apvnsid(N)$ to new physics
can be seen by rewriting $\qpv$ as
\begin{equation}
\qpv = \qw0\left[1+\delta_N\right]\ \ \ ,
\end{equation}
where
\begin{eqnarray}
\label{eq:deltan}
\delta_N&=&(Z\Delta\qwp+N\Delta\qwn)/\qw0\approx
-\zeta\left[\left({2Z+N\over N}\right)\hvu
	+\left({2N+Z\over N}\right)\hvd\right]\\
&=&-\zeta\left[(Z/N)(2\hvu+\hvd)
	+(2\hvd+\hvu)\right]\ \ \ \nonumber,
\end{eqnarray}
where the approximation $\qw0\approx -N$ has been made in light of the
small value for $1-4\sstw\approx 0.1$. From Eq. (\ref{eq:deltan})
we observe that for
atoms having $Z\approx N$, the weak charge is roughly equally sensitive to
the new up- and down-quark vector current interactions.

	The use of \lq\lq isotope ratios" involving $\apvnsid(N)$ and
$\apvnsid(N')$ largely eliminates the dependence on the atomic
structure-dependent
constant $\xi$ and the associated atomic theory uncertainty. We consider
two such ratios:
\begin{equation}
{\cal R}_1 = {\apvnsid(N')-\apvnsid(N)\over\apvnsid(N')+\apvnsid(N)}
\end{equation}
and
\begin{equation}
{\cal R}_2 = {\apvnsid(N')\over\apvnsid(N)}\ \ \ .
\end{equation}
To the extent that $\xi$ does not vary appreciably along the isotope chain,
one has
\begin{eqnarray}
{\cal R}_1 & = & {\qpv(N')-\qpv(N)\over \qpv(N')+\qpv(N)} \\
{\cal R}_2 & = & {\qpv(N')\over \qpv(N)} \ \ \ .
\end{eqnarray}
It is straightforward to work out the sensitivity of these ratios to
new physics. To this end, we write
\begin{equation}
{\cal R}_i = {\cal R}_i^0(1+\delta_i)\ \ \ ,
\end{equation}
where
\begin{eqnarray}
{\cal R}_1^0 & = & {\qw0(N')-\qw0(N)\over \qw0(N')+\qw0(N)} \\
{\cal R}_2^0 & = & {\qw0(N')\over \qw0(N)} \ \ \ ,
\end{eqnarray}
give the ratios in the Standard Model and the $\delta_i$ give corrections
arising from new physics. Letting $N'= N+\Delta N$ and dropping small
contributions containing $1-4\sstw$ one has
\begin{eqnarray}
{\cal R}_1^0 & \approx & {\Delta N\over 2N} \\
{\cal R}_2^0 & \approx & 1+{\Delta N\over N}
\end{eqnarray}
and
\begin{eqnarray}
\label{eq:deltaone}
\delta_1 &\approx &  \zeta\left({2Z\over N+N'}
	\right)(2\hvu+\hvd)\\
\label{eq:deltatwo}
\delta_2 &\approx & \zeta \left({Z\over N}\right)
	\left({\Delta N\over N'}\right)(2\hvu+\hvd) \ \ \ .
\end{eqnarray}

At first glance, the dependence of the $\delta_i$ $i=1,2$ on $\Delta\qwp=
\zeta(2\hvu+\hvd)$ only and not on $\Delta\qwn=\zeta(2\hvd+\hvu)$ may seem
puzzling. To
first order in $\zeta$, however, the shifts $\Delta\qwn$ appearing in the
numerator and
denominator of each ${\cal R}_i$ cancel. In the case of ${\cal R}_1$, for
example, one has
\begin{eqnarray}
\qpv(N')-\qpv(N)&\approx& -N'+N + (N'-N)\Delta\qwn\\
&=&(N-N')\left[1-\Delta\qwn\right]
\end{eqnarray}
and
\begin{eqnarray}
\qpv(N')+\qpv(N)&\approx&-(N+N') + (N+N')\Delta\qwn+2Z\Delta\qwp\\
&=&-(N+N')\left[1-\Delta\qwn-\left({2Z\over N+N'}\right)\Delta\qwp\right]
\end{eqnarray}
so that in the ratio, the dependence on $\Delta\qwn$ cancels to first
order. Hence,
the ${\cal R}_i$ are twice as sensitive to new physics involving $u$-quarks
than to
new physics which couples to $d$-quarks. The weak charge of a single
isotope, on the
other hand, has essentially the same sensitivity to $u$- and $d$-quark new
physics.

From a comparison of $\delta_N$ with the $\delta_i$, we also observe that, for
a given experimental precision, the
isotope ratios are generally less sensitive to direct new physics than is
the weak
charge for a
single isotope. This feature is particularly evident in the case of
${\cal R}_2$, since $\delta_2$ contains the explicit factor $\Delta N/N'$.
Taking $Z\approx N$ for the case of ${\cal R}_1$, we find that a single
isotope is three times more sensitive to new physics which couples to
$d$-quarks and 1.5 times more sensitive to the $u$-quark coupling. For new
physics scenarios which favor new $e-d$ interactions over $e-u$ interactions
({\em e.g.} $E_6$ models, discussed below), the weak charge for a single
isotope consititutes a more sensitive probe.

An alternative method for obtaining $\qpv$ is to scatter longitudinally
polarized electrons from fixed targets. Flipping the incident electron
helicity and comparing the helicity difference cross section with the
total cross section filters out the PV part of the weak neutral current
interaction. The resulting left-right asymmetry for elastic scattering
has the general form \cite{Mus94,Fei75,Wal77}
\begin{eqnarray}
\label{eq:alr}
\alr&=&{N_{+} - N_{-}\over N_{+} + N_{-}} \approx {2 M_\sst{NC}^\sst{PV}
\over M_\sst{EM}} \\
	&=& {G_F |q^2|\over 4\sqrt{2}\pi\alpha}\left[{\qpv\over Q_{EM}}
	+F(q)\right]\ \ \ .\nonumber
\end{eqnarray}
Here, $N_{+}$ ($N_{-}$) are the number of detected electrons for a positive
(negative) helicity incident beam; $M_\sst{EM}$ and $M_\sst{NC}^\sst{PV}$
are, respectively, the electromagnetic and parity-violating neutral current
electron-nucleus scattering amplitudes; $Q_{EM}$ is the nuclear EM charge;
and $F(q)$ is a correction involving hadronic and nuclear form factors. In
general, the latter term can be separated from the term containing the charges
by varying electron energy and angle. For elastic scattering, the weak charge
term can be isolated by going to forward angles and low energies. In the
case of PV M\" oller scattering, one has $F(q)\equiv 0$. The present PV
electron scattering program at MIT-Bates, Mainz-MAMI, and the Jefferson
Laboratory seeks to determine the $F(q)$ for a variety of targets, with a
special emphasis on contributions from strange quarks.

In order to compare the sensitivities of different scattering experiments
to new physics, we specify the terms in Eq. (\ref{eq:alr}) for the
following processes:
elastic scattering from the proton, $\alr(^1H)$; elastic scattering from
$(J^\pi, T)=(0^+,0)$nuclei , $\alr(0^+,0)$; excitation of the $\Delta(1232)$
resonance, $\alr(N\to\Delta)$; and M\" oller scattering, $\alr(e)$.
The corresponding charge terms are (neglecting Standard Model radiative
corrections)
\begin{eqnarray}
\qpv(^1H)/Q_{EM}(^1H) & = & (1-4\sstw)\left[1+\delp\right] \\
\qpv(0^+,0)/Q_{EM}(0^+,0) &=& -4\sstw\left[1+\delzp\right] \\
\qpv(e)/Q_{EM}(e) &=&(-1+4\sstw)\left[1+\dele\right]\ \ \ ,
\end{eqnarray}
while for the $N\to\Delta$ transition one replaces the ratio of
charges by the ratio of isovector weak neutral current and EM couplings:
\begin{eqnarray}
\qpv(N\to\Delta)/Q_{EM}(N\to\Delta)&\longrightarrow &2(1-2\sstw)
\left[1+\deld\right]\ \ \ .
\end{eqnarray}
The new physics corrections $\delta$ are given by
\begin{eqnarray}
\delp &=&\zeta(2\hvu+\hvd)/(1-4\sstw)\\
\delzp &=& -3\zeta(\hvu+\hvd)/(4\sstw)\\
\dele &=&-\zeta\hve/(1-4\sstw)\\
\deld &=&\zeta(\hvu-\hvd)/[2(1-2\sstw)]\ \ \ .
\end{eqnarray}

For completeness, we also write down the corresponding expressions for PV deep
inelastic scattering (DIS). We consider only the case of deuterium, which
was the
target in the first PV scattering experiment and was proposed in the early
1990's as the
target for a new SLAC experiment \cite{Bos92,Bos93}. An analysis of new
physics contributions to
the PV DIS asymmetry requires that we consider the more general four fermion
Lagrangian:
\begin{equation}
{\cal L}^\sst{NEW}={4\pi\kappa^2\over\Lambda^2} \sum_{q,i,j} h^q_{ij}\
{\bar e}_i
\gamma_\mu e_i \ {\bar q}_j \gamma^\mu q_j\ \ \ ,
\end{equation}
where \lq\lq i" and \lq\lq j" denote the handedness of the given fermion.
The $\hvq$
of Eq. (\ref{eq:lnew}) represent one linear combination of the $h_{ij}^q$:
\begin{equation}
\hvq=\left( h^q_{RR}-h^q_{LL}+h^q_{RL}-h^q_{LR}\right)/4 \ \ \ .
\end{equation}
The PV DIS asymmetry for a deuterium target is \cite{Mus94}
\begin{equation}
A_\sst{LR}^\sst{DIS}(^2{\hbox{H}}) = {G_F |q^2|\over
4\sqrt{2}\pi\alpha}{9\over 5}
\left\{{\tilde a}_1+{\tilde a}_2\left[{1-(1-y)^2\over
1+(1-y)^2}\right]\right\}\ \ \ ,
\end{equation}
where
\begin{eqnarray}
{\tilde a}_1&=& \left(1-{20\over
9}\sstw\right)\left[1+{\tilde\delta}_1\right]\\
{\tilde a}_2&=& \left(1-4\sstw\right)\left[1+{\tilde\delta}_2\right]\ \ \ .
\end{eqnarray}
The ${\tilde\delta}_i$ contain Standard Model radiative corrections,
corrections involving
the quark distribution functions \cite{Mus94}, and contributions from new
physics. Writing only
the latter, we obtain
\begin{eqnarray}
{\tilde\delta}_1&=& {\zeta\over24(1-20\sstw/9)}\ (2\hvu-\hvd)\approx
\zeta(2\hvu-\hvd)/12\\
{\tilde\delta}_2&=& {\zeta\over 2(1-4\sstw)}\ \sum_q
Q_q\left[h^q_{RR}-h^q_{LL}+h^q_{LR}-
	h^q_{RL}\right]\\
&\approx& 6.5\zeta\ \sum_q Q_q\left[h^q_{RR}-h^q_{LL}+h^q_{LR}-
	h^q_{RL}\right]\ \ \ ,
\end{eqnarray}
where $Q_q$ is the quark EM charge.

The expressions for the various $\delta_i$ allow us to make a few observations
regarding the relative sensitivities the corresponding observables to new
physics.
For this purpose, we take $\hvu=\hvd=1$ and specify $\delta_N$ for the case of
$^{133}$Cs. We also use cesium for the isotope ratios and take a reasonable
range of neutron numbers:$N=75$, $N'=95$ \cite{Mas95}. In Table I we show the
$\delta_i$ in units of $\zeta$. The third column gives a scale factor $f$
defined as
\begin{equation}
f_i=\sqrt{\delta_i/\delta_N}\ \ \ .
\end{equation}
The factor $f_i$ can be used to scale the cesium APV sensitivity to the new
physics
mass scale $\Lambda$ to those obtainable from any other observable when
measured
with the same precision as $\qpv({\hbox{Cs}})$:
$\Lambda(i)=f_i\Lambda({\hbox{Cs}})$.
Alternatively, the sensitivity of any other observable will be the same as
that
of cesium when the precision is $f_i^2$ times the cesium uncertainty.
The numbers shown in the Table are obtained using the
$\overline{\hbox{MS}}$ value
$\sstw=0.2314$ \cite{PDG98} in tree-level expressions for the weak charges.
The entries for the M\" oller asymmetry have been modified to account for
one-loop
electroweak radiative corrections, according to the calculation of Ref.
\cite{Cza96}.
In the latter case, these corrections reduce the asymmetry by $\approx$
40\% from its
tree-level value. Radiative corrections do not appreciably alter the
relative new physics
sensitivities of the other observables listed in Table I.

$$\hbox{\vbox{\offinterlineskip
\def\strut{\hbox{\vrule height 15pt depth 10pt width 0pt}}
\hrule
\halign{
\strut\vrule#\tabskip 0.2cm&
\hfil$#$\hfil&
\vrule#&
\hfil$#$\hfil&
\vrule#&
\hfil$#$\hfil&
\vrule#&
\hfil$#$\hfil&
\vrule#\tabskip 0.0in\cr
& \multispan3{\hfil\bf TABLE I\hfil} & \cr\noalign{\hrule}
& \hbox{Correction}\quad \delta&& \hbox{Scale factor}\quad f_i
& \cr\noalign{\hrule}
& \delta_N\approx 5.1\zeta && 1  & \cr
& \delta_1\approx 1.9\zeta && 0.6 &\cr
& \delta_2\approx 0.4\zeta && 0.3 &\cr
& \delp\approx 40\zeta && 2.8 &\cr
& \delzp\approx 6.5\zeta && 1.1 &\cr
& \dele\approx 22\zeta && 2.1 & \cr
& \deld = 0 && 0 &\cr
& \tilde\delta_1\approx\zeta/12 && 0.13 &\cr
\noalign{\hrule}}}}$$
{\noindent\narrower {\bf Table I.} \quad Relative sensitivities of PV
observables
to new physics, assuming $\hvu=\hvd$, tree-level values for the corresponding
weak charges (except for the M\" oller asymmetry, as noted in the text), and
$\sin^2\hat{\theta}_\sst{W}=0.2314$. The scale factor
$f_i=\sqrt{\delta_i/\delta_N}$ can be used to scale mass bounds from the
cesium APV bounds to the bounds for observable $i$ assuming the same
precision for
both $\delta_N$ and $\delta_i$. Note that we have assumed $\hvu=\hvd$ so that
$\deld=0$.
\smallskip}

As Table I illustrates, $\alr(^1{\hbox{H}})$ and $\alr(e)$ display the
greatest sensitivities to
new physics for a given level of error in the observables. The reason is the
suppression of $Q_W^0$ for the proton and electron, which goes as
$(1-4\sstw)$ at tree level,
as well as the additional suppression of $Q_\sst{W}^e$ due to radiative
corrections.
This suppression, however, renders the attainment of high precision more
difficult
than for some of the other cases, since the statistical uncertainty in $\alr$
goes as $1/\alr$ \cite{Mus92,Mus94}. To set the scale, we note that a 10\%
$\alr(^1{\hbox{H}})$
measurement would be as sensitive as the present cesium APV determination
to the mass scale
$\Lambda$. Given
the performance of the beam and detectors at the Jefferson Lab, it appears
that a
future measurement of $\alr(^1{\hbox{H}})$ with  5\% or better precision
could be feasible
\cite{Sou98}. Such a determination would yield new physics limits
comparable to those
from cesium APV should the atomic theory error  be reduced to the level of
the present
experimental error. A 2.5\% $ep$ measurement would strengthen the present
APV bounds
by a factor of two. Sensitivity at this level would be competitive with those
expected from high energy colliders by the end of the next decade
\cite{Riz96a,Riz96b}.
The physics reach of a
6\% determination of the M\" oller asymmetry would be similar to that of
the present cesium
measurement, though PV $ee$ scattering is in general sensitive to a
different set of new
interactions than arise in the $eq$ sector.

In the case of isotope ratios, which depend like $\alr(^1{\hbox{H}})$ on
$\Delta\qwp$, a 0.5\%
determination ${\cal R}_1$ would give new physics limits comparable to the
present cesium
results. The prospects for achieving this precision are promising. The
Berkeley group, for
example, expects to perform a 0.1\% determination of ${\cal R}_1$ using the
isotopes of Yb
$N=100\to N=106$
\cite{Bud98}.\footnote{Nuclear structure uncertainties may cloud the
interpretation of
such a measurement, however (see Section V)}. A measurement of such
precision would
double the present cesium sensitivity, neglecting nuclear structure
corrections. Similarly, the
Seattle group plans to conduct studies on the isotopes of Ba$^+$ ions
\cite{For97}. For both Yb and Ba, the scale factors $f_1$ are similar to
that for cesium isotopes,
whereas $f_2$ depends strongly on the range $\Delta N$.

As the discussion of the following section illustrates, variations from
this general pattern
of relative sensitivities occur when specific new physics scenarios are
considered. For
example, our assumption of purely isoscalar new interactions ($\hvu=\hvd$) in
arriving at Table I renders the PV $N\to\Delta$ correction zero. In the
case of purely
isovector interactions, the scale factor for PV $ep$ scattering becomes 6.6
while that for
the PV $N\to\Delta$ asymmetry is 2.5. In short, the weak charge for a
single heavy isotope is
relatively insensitive to new isovector interactions. As a second example, the
M\" oller asymmetry is at least an order of magnitude less sensitive to
leptoquarks than are the other observables, even though it generally
displays a relatively
strong sensitivity to new heavy physics (see discussion in Section IV).
Similarly, in $E_6$
models which give rise to leptoquarks, one has $\hvu=0$ while $\hvd\not=0$.
In this case,
systems having a relatively large
$d$-quark to $u$-quark ratio are advantageous. The scale factor $f$ for PV $ep$
scattering, for example, is reduced to 2.2 when considering such $E_6$
models. A similar
reduction occurs in the scale factors for the isotope ratios ${\cal R}_i$,
since
these ratios, like $\alr(^1{\hbox{H}})$, are sensitive primarily to
$\Delta\qwp$. We also note in
passing that limits from high energy colliders are sometimes quoted
assuming that the
new physics couplings to $u$- and $d$-quarks are the same as in the
Standard Model. While
there is no {\em a priori} reason to invoke this assumption, it would imply
that the
new physics shifts $\delp$ and $\delta_i$ ($i=1,2$) are suppressed by the
same $1-4\sstw$
factor which enters $\qwp$ at tree level.

Finally, we make a few observations regarding the new physics corrections
to the DIS
asymmetry. The correction ${\tilde\delta}_1$ depends on the same
combination of the
$h_{ij}^q$ that arises in the other PV observables, but with a different
$u$- and
$d$-quark weighting than appears anywhere else. As reflected in Table I,
however,
the sensitivity of ${\tilde\delta}_1$ to new physics is much weaker than
for most
of the other observables. The correction ${\tilde\delta}_2$, on the other
hand, is
significantly more sensitive to new four fermion interactions than is
${\tilde\delta}_1$.
Moreover, its dependence on the $h_{ij}^q$ differs from that of all the
other PV observables
discussed here. In fact, certain scenarios proposed for evading the atomic
PV limits
on the $h_{ij}^q$, such as SU(12) symmetry \cite{Nel97}, would not apply to
bounds of comparable
strength obtained from ${\tilde\delta}_2$. Unfortunately, a precise
determination of the ${\tilde
a}_2$ term in the DIS asymmetry appears to be  difficult.

\section{Model Illustrations}
\label{sec:model}

The interaction of Eq. (\ref{eq:lnew}) may be specified for different new
physics scenarios.
In what follows, we consider three examples which illustrate
the relative sensitivies of PV observables to different models:
(a) additional neutral gauge bosons,  (b)
lepto-quarks and $R$-parity violating supersymmetric models, and (c)
fermion compositeness.

\medskip

\noindent {\bf A. Additional neutral gauge bosons.}
\medskip

The existence of
additional, neutral gauge bosons is natural in the context of
superstring-inspired E$_6$ theories, in which the spontaneous breakdown of
E$_6$ symmetry results in the existence of one or more $U(1)$ gauge
symmetries beyond the U(1)$_Y$ of the Standard Model
\cite{Lon86,Lan92,Lan95,Moh92}.
Additional neutral gauge bosons may also arise in left-right symmetric
models \cite{Lan92,Moh92}.
It is conceivable that at least one of the neutral gauge bosons is sufficiently
light to be of interest to low-energy neutral current processes. We let
$Z'$ and $Z$ denote the \lq\lq new" and Standard Model neutral gauge bosons,
respectively. The exisentence of a light $Z'$  which mixes with the $Z$
is ruled out by $Z$-pole observables. In the event that the $Z-Z'$ mixing
angle is $\approx 0$, however, LEP and SLC measurements provide rather
weak constraints \cite{Lan95}. Consequently, we consider the case of zero
mixing.

For the sake of illustration, we follow the E$_6$ analysis of Ref.
\cite{Lon86},
in which the different symmetry breaking scenarious can be parameterized
by writing the $Z'$ as
\begin{equation}
\label{eq:zprime}
Z'= \cos\phi Z_\psi + \sin\phi Z_\chi \ \ \ .
\end{equation}
The $Z_\psi$ and $Z_\chi$ arise, for example, from the breakdown
E$_6\to$ SO(10)$\times$ U(1)$_\psi$ and SO(10)$\to$ SU(5)$\times $ U(1)$_\chi$.
Since the multiplets of SO(10) contain both $f$ and $\bar{f}$ for the
leptons and quarks of the Standard Model, $C$-invariance implies that the
$Z_\psi$ can have only axial vector couplings to these fermions. As a result,
it cannot contribute at tree-level to low-energy PV observables. In the case
of SU(5), however, the left-handed $d$-quark and $e^+$ live in a
different multiplet from the left-handed $\bar{d}$ and $e^-$, whereas
the $u$ and $\bar{u}$ live in the same multiplet. The $Z_\chi$ correspondingly
has both vector and axial vector couplings to the electron and $d$-quarks,
and only axial vector $u$-quark couplings. In short, E$_6$ $Z'$ bosons
yield $\hvu=0$ and $\hvd$, $\hve$ $\propto\sin\phi$.

According to the notation of Eq. (\ref{eq:lnew}), we have for E$_6$ models
\begin{eqnarray}
\kappa^2&=&\alpha' \\
\Lambda^2&=& M_{Z'}^2\\
\hvu&=&0 \\
\hvd=-\hve&=&\left[\sin^2\phi-\sqrt{15}\sin\phi\cos\phi/3\right]/20\ \ \ ,
\end{eqnarray}
where $\alpha'$ is the fine structure constant associated with the new
gauge coupling. Generally, one has\cite{Lan92}
\begin{equation}
\label{eq:alphap}
\alpha'\simle{5\over 3} {\alpha\over\cos^2\theta_\sst{W}}\approx 2.2\alpha
\ \ \ .
\end{equation}

Different models for the $Z'$ correspond to different choices for
$\phi$. Examples include the $Z_\eta$ ($\tan\phi=-\sqrt{3/5}$) and
the $Z_I$ ($\tan\phi=-\sqrt{5/3}$), where the latter is associated
with an additional \lq\lq inert" SU(2) gauge group not contributing to
the electromagnetic charge. From the
standpoint of phenomenology, it is  worth noting the dependence of $\hvd$
and $\hve$ on the value
of $\phi$. For $\phi=\phi_c=\tan^{-1}(\sqrt{5/3})\approx 52^\circ$,
$\hvd=0=\hve$. For
$\phi>\phi_c$,
$\hvd>0$. From Eq. (\ref{eq:deltan}), we observe that $\delta_N$ is
negative for
$\hvu=0$ and $\hvd>0$. The most recent value of  $\delta_N$ for cesium
implies that $\hvd>0$ at the one $\sigma$ level, and therefore could not be
explained models
giving $\phi<\phi_c$. A model which gives nearly the largest possible
contribution to
the weak charge is the $Z_\chi$, which corresponds to $\phi=90^\circ$.

An interesting variation on the idea of extended gauge group symmetry is
that of left-right symmetric theories. In such theories, the low energy
gauge group becomes SU(2$)_L\times$SU(2$)_R\times$U(1$)_{B-L}$, where
$B-L=1/3$ for baryons and $-1$ for leptons. In the case of \lq\lq manifest"
left-right symmetry the SU(2$)_L$ and SU(2$)_R$ couplings are identical.
For this case, a second low-mass neutral gauge boson $Z_{R}$ couples to
fermions with the strengths \cite{Moh92}
\begin{eqnarray}
\hvu&=&-\frac{3}{5}\frac{\alpha}{4}\left(\frac{\alpha}{4}-\frac{1}{6\alpha}
\right)\\
\hvd&=&\frac{3}{5}\frac{\alpha}{4}\left(\frac{\alpha}{4}+\frac{1}{6\alpha}
\right)\\
\hve&=&\frac{3}{5}\frac{\alpha}{4}\left(\frac{\alpha}{4}-\frac{1}{2\alpha}
\right)
\end{eqnarray}
where 
\begin{equation}
\alpha=\left({1-2\sstw\over\sstw}\right)^{1/2}\approx 1.53\ \ \ .
\end{equation}
With this set of couplings, the combination appearing in the correction
$\Delta\qwp$
is $2\hvu+\hvd\approx 0.012 << \hvu, \ \hvd$. Consequently,
the sensitivies of the $R_i$ and $\alr(^1{\hbox{H}})$ are suppressed relative
to their generic scale. The corresponding mass limits on $M_{Z_{R}}$ are weaker
than those obtainable from cesium APV or $\alr(0^+,0)$.

In Table II, we give the present and prospective sensitivities for two
species of additional
neutral gauge bosons, the $Z_\chi$ and $Z_{R}$. In particular,
we show lower bounds on the Fermi constant associated
with the new gauge boson $Z'$, defined as
\begin{equation}
{G_a^\prime\over \sqrt{2}} \equiv {g^{\prime\ 2}\over 8 M_{Z^\prime_a}^2}\
\ \ ,
\end{equation}
where $g^\prime$ is the coupling associated with the additional $U(1)_a$
gauge group.
Low-energy PV observables constrain the ratio $(g^\prime/M_{Z^\prime_a})$
and do not provide
separate limits on the mass and coupling. Consequently, the ratio of
$G^\prime_\chi/G_F$
characterizes the strength of a new $U(1)_\chi$ gauge interaction relative
to the strength
of the Standard Model. In general, mass bounds for the $Z'$ can be obtained
from the limits on
$G'$ under specific assumptions for $g'$. A comparison of such mass bounds
is often
instructive, so we quote such bounds in the final two columns of Table II.
Lower bounds on $M_\chi$ are quoted assuming the
maximal value for $g'$ as given by Eq. (\ref{eq:alphap}). In the case of LR
symmetry models with
manifest LR symmetry, one has $g'=g$. The corresponding mass limits for the
$Z_{LR}$ are given
in the final column of Table II. Since we only discuss the case of manifest
LR symmetry
above, we do not include bounds on $G^\prime_{LR}/G_F$.

$$\hbox{\vbox{\offinterlineskip
\def\strut{\hbox{\vrule height 15pt depth 10pt width 0pt}}
\hrule
\halign{
\strut\vrule#\tabskip 0.2cm&
\hfil$#$\hfil&
\vrule#&
\hfil$#$\hfil&
\vrule#&
\hfil$#$\hfil&
\vrule#&
\hfil$#$\hfil&
\vrule#&
\hfil$#$\hfil&
\vrule#\tabskip 0.0in\cr
& \multispan9{\hfil\bf TABLE II\hfil} & \cr\noalign{\hrule}
& \hbox{Observable} && \hbox{Precision} && G^\prime_\chi/G_F && M_{Z_\chi}\
({\hbox{GeV}})&&
M_{Z_{LR}}
\ ({\hbox{GeV}})
& \cr\noalign{\hrule}
& \qpv({\hbox{Cs}}) && 1.3\% && 0.006 && 730 && 790& \cr
&  && 0.35\%  && 0.0016 && 1410 && 1520 &\cr
& {\cal R}_1 && 0.3\% && 0.006 && 740 && 360 & \cr
&  && 0.1\% && 0.002 && 1300 && 630 & \cr
& \qpv(^1{\hbox{H}})/Q_\sst{EM}(^1{\hbox{H}}) && 10\% && 0.010 && 580 &&
285 & \cr
& && 3\% && 0.003 && 1100 && 520 & \cr
& \qpv(0^+,0)/Q_\sst{EM}(0^+,0) && 1\% && 0.004 && 910 && 920 & \cr
& \qpv(e)/Q_\sst{EM}(e) && 7\% && 0.004 && 910 && 460 & \cr
& \alr(N\to\Delta) && 1\% && 0.013 && 490 && 920 & \cr
&  {\tilde a}_1 && 1\% && 0.15 && 145 && 320 &\cr
\noalign{\hrule}}}}$$
{\noindent\narrower {\bf Table II.} \quad Present and prosepctive limits on
two species of additional neutral gauge bosons. The third column gives the
ratio
of fermi constants as defined in the text. The fourth and fifth columns give
lower bounds on masses for the $Z_\chi$ and $Z_{LR}$, respectively,
assuming the
precision given in column two.
\smallskip}

The limits in Table II lead to several observations. Primary among these is
that
low-energy PV already constrains the strength of new, low-energy gauge
interactions
to be at most a few parts in a thousand relative to the strength of the
SU(2$)_L\times$U(1$)_Y$ sector. When reasonable assumptions are made about
new gauge
couplings strengths, low-energy mass bounds now approach one TeV. The
significance of
these bounds becomes more apparent when a comparison is made with the
results of collider
experiments. The present 110 pb$^{-1}$ $p\bar p$ data set analyzed by the
CDF collaboration
yields a lower bound on $M_{Z_{LR}}$ of 620 GeV, assuming manifest LR
symmetry \cite{Abe97a}. The
lower bound for $M_{Z_\chi}$ is 585 GeV, assuming no $Z_\chi$ decays to
supersymmetric
particles\cite{Abe97a}. The sensitivity of cesium APV already exceeds these
Tevatron bounds.
In fact, collider experiments and low-energy PV provide complementary
probes of extended gauge
group structure. PV observables are sensitive to the vector couplings of
the $Z'$ to fermions.
For a model for which this coupling is small or vanishing ({\em e.g.}, the
$Z_\psi$ having
$\phi=0^\circ$ in Eq. (\ref{eq:zprime}), PV observables cannot yield
significant information.
Collider experiments, on the other hand, retain a sensitivity to such $Z'$
interactions. For
models in which the $ffZ'$ coupling is not suppressed, low-energy PV
presently displays the
greatest sensitivity.

A look to the future suggests that PV could continue to play such a
complementary role.
Assuming the collection of 10 fb$^{-1}$ of data at TeV33, for
example, the current Tevatron bounds on $M_{Z'}$ would increase by roughly
a factor of
two \cite{Riz96a}. The prospective sensitivity of cesium APV, assuming a
reduction in atomic
theory error to the level of the present experimental uncertainty, would
exceed the collider
reach by $\sim$ 50\%. Precise determinations of the isotope ratio ${\cal
R}_1$ or various PV
electron scattering asymmetries could also yield sensitivities which match
or exceed the
prospective TeV33 bounds. Only with the advent of the LHC or $\simge 60$
TeV hadron collider will
high-energy machines probe masses significantly beyond those accessible
with low-energy PV
\cite{Riz96a}.

Finally, Table II illustrates the model-sensitivity of different PV
observables. For the
models considered here, the mass bounds do not scale with the $f_i$ of
Table I since
$\hvu\not=\hvd$. Both the $Z'$ in $E_6$ and the $Z_{LR}$ couple more
strongly to neutrons
than protons. Consequently, both $R_1$ and $\alr(^1{\hbox{H}})$ display
weaker sensitivity
to new gauge interactions than their generic sensitivies to new physics
indicated in Table I.

\bigskip
\noindent{\bf B. Leptoquarks and Supersymmetry}

\medskip
In early 1997, the H1 \cite{Adl97} and ZEUS \cite{Bre97} collaborations
reported the presence
of anomalous events in high-$|q^2|$ $e^+p$ collisions at HERA. These
events have been widely interpreted as arising from s-channel lepton-quark
resonances with mass $\mlq \approx 200$ GeV \cite{Hew97,LQ97}. Given the
stringent limits
on the existence of vector leptoquarks (LQ's) obtained at Fermilab
\cite{Hew97,Hob97,Alt97},
scalar leptoquarks are the favored interpretation of the HERA events. Although
the results remain controversial,
they are nonetheless provocative and suggest a consideration of LQ effects
in low-energy PV processes. To that end, we consider general LQ interactions
of the form
\begin{eqnarray}
\label{eq:leptoa}
{\cal L}_\sst{LQ}^\sst{S}&=& \lscal(\phi{\bar e}_L q_R+{\hbox{h.c.}})\\
{\cal L}_\sst{LQ}^\sst{V}&=& \lvect({\bar e}_L\gamma_\mu q_L\phi^\mu
	+{\hbox{h.c.}})
\end{eqnarray}
where $\phi$ and $\phi^\mu$ denote scalar and vector LQ fields, respectively.
For simplicity, we do not explicitly consider the corresponding interactions
obtained from Eq.~(\ref{eq:leptoa}) with $L\leftrightarrow R$. The
corresponding analysis is
similar to what follows. Assuming $\mlqs >> |q^2|$, the process $eq\to\ LQ\
\to eq$
gives rise to the following PV interactions:
\begin{eqnarray}
{\cal L}^\sst{S}_\sst{PV}&=&(\lscal/2\mlq)^2 \left[{\bar e}q {\bar q}\gamma_5 e
	-{\bar e}\gamma_5 q{\bar q} e\right]\\
{\cal L}^\sst{V}_\sst{PV}&=&(\lvect/2\mlq)^2\left[{\bar e}\gamma_\mu q{\bar q}
\gamma^\mu \gamma_5 e+{\bar e}\gamma_\mu \gamma_5 q{\bar q}\gamma^\mu e\right]
\end{eqnarray}
After a Fierz transformation, these become
\begin{eqnarray}
{\cal L}^\sst{S}_\sst{PV}&=&(\lscal/2\sqrt{2}\mlq)^2[{\bar e}\gamma_\mu
	\gamma_5 e{\bar q}\gamma^\mu q -{\bar e}\gamma_\mu e {\bar q}
	\gamma^\mu\gamma_5 q\\
	&& +{1\over 4} {\bar e}\sigma_{\mu\nu} e {\bar q}\sigma^{\mu\nu}
	\gamma_5 q - {1\over 4}{\bar e}\sigma_{\mu\nu}\gamma_5 e
	{\bar q}\sigma_{\mu\nu} q]\nonumber \\
&& \nonumber \\
{\cal L}^\sst{V}_\sst{PV}&=&-(\lvect/2\mlq)^2[{\bar e}\gamma_\mu e
	{\bar q}\gamma^\mu\gamma_5 q +{\bar e}\gamma_\mu\gamma_5 e
	{\bar q}\gamma^\mu q]
\end{eqnarray}
In terms of the interaction in Eq. (\ref{eq:lnew}), we may identify
\begin{eqnarray}
\Lambda^2&=&\mlqs\\
\kappa^2&=&\lambda^2/16\pi
\end{eqnarray}
and $\hvq=1/2$ ($\hvq=-1$) for scalar (vector) LQ interactions.

Assuming for simplicity that either a u-type or d-type LQ (but not
both) contributes to low-energy PV processes, the results from cesium
APV, together with Eqs. (\ref{eq:lnew}) and (\ref{eq:deltan}), yield the
following $1\sigma$
limits on LQ couplings and masses:
\begin{equation}
\label{eq:leptoscal}
\lscal\leq\cases{0.042\  (\mlq/100\ \mbox{GeV})\ , & u-type \cr
&\cr
0.04\ (\mlq/100\ \mbox{GeV})\ , & d-type\cr}
\end{equation}
and
\begin{equation}
\lvect\leq\cases{0.030 (\mlq/100\ \mbox{GeV}), & u-type\cr
&\cr
0.028(\mlq/100\ \mbox{GeV}), & d-type\cr} \ \ \ .
\end{equation}

Substituting the HERA value of $\mlq\approx$ 200 GeV into Eq.
(\ref{eq:leptoscal}) yields an
upper bound of $\lscal\leq 0.08$. On general grounds, one might have
expected $\kappa^2\sim\alpha$ or $\lscal\sim 0.6$. The cesium APV results
require the coupling for a 200 GeV scalar LQ to be about an order of
magnitude smaller than this expectation. Alternatively, if one does
not interpret the HERA results as a 200 GeV LQ and assumes $\kappa^2
\sim\alpha$, the APV bounds on the scalar LQ mass are $\mlq>$ 1.5 TeV.
These LQ constraints
are consistent with those obtained from high-energy collider experiments,
though
low- and high-energy processes generally provide complementary information.
The constraints
from the Tevatron\cite{Abb98}, for example, are essentially
$\lscal$-independent, while providing
bounds on $\mlq$ and LQ decay branching fraction \cite{Hew97,Hew98}.

Table III gives comparable bounds on the LQ coupling-to-mass ratio for
the other PV observables discussed in Section III. The bounds are characterized
by the quantity $\gamma_q$, defined as
\begin{equation}
\label{eq:leptob}
\lscal \leq \gamma_q\ (\mlq/100\ \mbox{GeV}) \ \ \ ,
\end{equation}
where $q$ denotes the quark flavor.

$$\hbox{\vbox{\offinterlineskip
\def\strut{\hbox{\vrule height 15pt depth 10pt width 0pt}}
\hrule
\halign{
\strut\vrule#\tabskip 0.2cm&
\hfil$#$\hfil&
\vrule#&
\hfil$#$\hfil&
\vrule#&
\hfil$#$\hfil&
\vrule#&
\hfil$#$\hfil&
\vrule#\tabskip 0.0in\cr
& \multispan7{\hfil\bf TABLE III\hfil} & \cr\noalign{\hrule}
& \hbox{Observable} && \hbox{Precision} && \gamma_u && \gamma_d
& \cr\noalign{\hrule}
& \qpv({\hbox{Cs}}) && 1.3\% && 0.04 && 0.042 & \cr
&  && 0.35\%  &&  0.021 && 0.022 &\cr
& {\cal R}_1 && 0.3\% && 0.04 && 0.028 & \cr
&  && 0.1\% && 0.023 && 0.016 & \cr
& \qpv(^1{\hbox{H}})/Q_\sst{EM}(^1{\hbox{H}}) && 10\% && 0.05 && 0.036 & \cr
& && 3\% && 0.028 && 0.02 & \cr
& \qpv(0^+,0)/Q_\sst{EM}(0^+,0) && 1\% && 0.033 && 0.033 & \cr
& \qpv(e)/Q_\sst{EM}(e) && 7\% && -- && -- & \cr
& \alr(N\to\Delta) && 1\% && 0.06 && 0.06 & \cr
&  {\tilde a}_1 && 1\% && 0.14 && 0.20 &\cr
\noalign{\hrule}}}}$$
{\noindent\narrower {\bf Table III.} \quad Present and prosepctive limits on
leptoquark interactions. Third and fourth columns give $\gamma_q$ for a
$q$-type leptoquark, as defined in Eq. (\ref{eq:leptob}). Leptoquark
sensitivity of
M\"oller asymmetry does not behave according to Eq. (\ref{eq:leptob}), so
that no limits
on the $\gamma_q$ are attainable.  \smallskip}

Note that no bounds
are given for the M\" oller asymmetry, as LQ's do not contribute at
tree level. The leading contributions arise from the loop graphs of
Fig 1. We have evaluated the amplitudes for these diagrams and obtain the
following
contributions to the PV effective $ee$ interaction (to leading order in
fermion masses and
momenta):
\begin{eqnarray}
{\cal L}^\sst{PV}_{(a)}&=& \left({\lambda_\sst{S}^2\over
16\pi\mlq}\right)^2{\bar e}\gamma_\mu e
	{\bar e}\gamma^\mu\gamma_5 e \\
{\cal L}^\sst{PV}_{(b)}&=& {\alpha Q_q\over
12\pi}\left({\lscal\over\mlq}\right)^2
	\ln {m_q\over\mlq} {\bar e}\gamma_\mu e{\bar e}\gamma^\mu
	\gamma_5 e
\end{eqnarray}
where $m_q$ and $Q_q$ are the intermediate state quark mass and
E.M. charge. For $\mlq=100$ GeV, a 7\% determination of the M\" oller
asymmetry would yield
\begin{equation}
\lscal\leq 1.06
\end{equation}
from graph (a) and
\begin{equation}
\lscal\leq
\cases{0.88, & d-type \cr
	      0.6, & u-type\cr}
\end{equation}
from graph (b). The limits for a vector LQ are comparable. The prospective
M\" oller bounds are more than an order of magnitude weaker than those
attainable with
semi-leptonic PV. Any deviation of the M\" oller asymmetry from the
Standard Model prediction is unlikely to be due to LQ's.

Scalar leptoquarks arise naturally in R-parity violating
supersymmetric theories from a term in the superpotential of the
form\cite{Bar89}
\begin{equation}
\lambda^\prime_{ijk}L^i_L Q^j_L {\bar D}^k_R
\end{equation}
where the chiral superfields $L_L^i$, $Q_L^j$, and ${\bar D}_R^k$ contain
the left-handed lepton and quark doublets and right handed d-quark
singlets, respectively, for generations $i,j,k$. This term includes
a lepton-number violating electron-quark-squark
interaction\cite{Bar89,Kon97,Cho97}
\begin{equation}
\label{eq:rpva}
{\cal L} = \lambda^\prime_{1jk}\left[ {\bar d}_{kR} e_L \phi^u_{jL}+
	{\bar e}_L d_{kR}{\bar\phi}^u_{jL}-{\bar e}^c u_{jL}
	{\bar\phi}^d_{kR}-{\bar u}_{jL}e^c\phi^d_{kR}\right]
\end{equation}
where $\phi^u_{jL}$ is the squark of charge $+2/3$ associated with
a left-handed $+2/3$ charged quark of generation $j$, etc. The first
two terms in Eq. (\ref{eq:rpva}) contribute to the HERA processes for $k=1$
when a positron scatters from a valence d-quark in the proton, while the
last two terms contribute for scattering from a sea u-quark. Low-energy
PV receives a contribution from both terms. For illustrative purposes, we
consider
only the first two. Identifying the
$\lambda^\prime_{1j1}$ with $\lscal$ of Eq. (\ref{eq:leptoa}), we obtain

\begin{equation}
\lambda^\prime_{1j1} \leq 0.04 (M_{\phi^u_{jL}}/100\ \mbox {GeV})
\end{equation}
as the bound obtained from cesium APV. The prospective bounds attainable
from other PV observables may be obtained from Table III\footnote{The most
stringent bounds on $\lambda^\prime_{111}$ are derived from neutrinoless double
$\beta$-decay\cite{Hir96}}.

For completeness, we note that low-energy PV is sensitive to another
R-breaking term
\begin{equation}
\lambda_{ijk}L_L^i L_L^j {\bar E}_R^k
\end{equation}
where ${\bar E}_R^k$ contains the right-handed charged-lepton singlet fields.
This term generates a four-fermion contact interaction which contributes
to $\mu$-decay \cite{Bar89}:
\begin{equation}
\label{eq:rpvb}
{\cal L} = -(\lambda_{12k}/\sqrt{2} M_{\phi^e_{kR}})^2 {\bar e}_L\gamma_\alpha
	\nu^e_L {\bar\nu}^\mu_L \gamma^\alpha \mu_L \ \ \ .
\end{equation}
Because the strength of the weak neutral current amplitude $(g/M_W)^2$
is written in terms of the $\mu$-decay Fermi constant, $G_\mu$, the
interaction (\ref{eq:rpvb}) induces a correction to low-energy PV interactions:
\begin{equation}
{g^2\over 8 M_W^2} = {G_\mu\over\sqrt{2}}-\left(\lambda_{12k}/2\sqrt{2}
	M_{\phi^e_{kR}}\right)^2\equiv{G_\mu\over\sqrt{2}}[1-\Delta_{12k}]
\end{equation}
where
\begin{equation}
\Delta_{12k} = {\lambda^2_{12k}\over 4\sqrt{2}G_\mu M^2_{\phi^e_{kR}}}
\end{equation}
A one percent determination of {\em any} low-energy PV observable (including
the M\" oller asymmetry) would yield the bounds
\begin{equation}
\lambda_{12k}\leq 0.08 (M_{\phi^e_{kR}}/100\ \mbox{GeV})
\end{equation}

It is instructive to compare this bound with that obtained from
superallowed $\beta$-decay. In the latter case, interaction (\ref{eq:rpvb})
would cause the measured value of the CKM matrix element $|V_{ud}|$
to differ from a valued implied by CKM matrix unitarity \cite{Bar89}. Letting
$|V_{ud}|_\sst{EX}$ denote the value extracted from experiment -- assuming
only the Standard Model -- and $|V_{ud}|$ the value implied by unitarity, one
has
\begin{equation}
\label{eq:vud}
|V_{ud}|^2_\sst{EX}=|V_{ud}|^2\left[1-2\Delta_{12k}\right]
\end{equation}

The experimental situation regarding  superallowed beta decay has
generated some debate about the value of $|V_{ud}|_\sst{EX}$. Assuming
the experimental values for $|V_{us}|$ and $|V_{ub}|$, one finds from
a fit to nine precisely measured superallowed ${\cal F}t$ values
\cite{Tow95,Hag96}
\begin{equation}
\label{eq:global}
|V_{ud}|^2_\sst{EX}-|V_{ud}|^2 = -0.0023 \pm 0.0013
\end{equation}
A recent measurement of the superallowed $^{10}$C beta decay, however,
yields a value consistent with CKM unitarity at the 1$\sigma$ level
\cite{Fuj98}:
\begin{equation}
|V_{ud}|^2_\sst{EX}-|V_{ud}|^2 = -0.001 \pm 0.0027
\end{equation}

The $^{10}$C result, together with Eqn. (\ref{eq:vud}), requires that
\begin{equation}
\label{eq:carbon}
2\Delta_{12k} |V_{ud}|^2 \leq 0.0027
\end{equation}
or
\begin{equation}
\label{eq:betad}
\lambda_{12k}\leq 0.03 (M_{\phi^e_{kR}}/100\ \mbox{GeV})
\end{equation}
If, on the other hand, one assumes that the $2\sigma$ deviation is due to
some type of new physics, then it could  be generated by the lepton number
violating interaction (\ref{eq:rpvb}), since $\Delta_{12k}$ enters Eq.
(\ref{eq:vud}) with the
correct sign\footnote{In general leptoquark models where both u- and d-type
LQ's
contribute to $\beta$-decay, the sign of the corresponding correction to
unitarity may
also be consistent with Eq. (\ref{eq:global}) under certain
assumptions\cite{Hew98}.}.  In this
case, the inequalities in Eqs. (\ref{eq:carbon}) and (\ref{eq:betad}) would
be replaced by
the appropriate equalities.

\medskip
\noindent{\bf C. Compositeness}
\medskip

The Standard Model assumes the known bosons and fermions to be pointlike.
The possibility that they possess internal structure, however, remains an
intriguing one. Manifestations of such composite structure could include
the presence of fermion form factors in elementary scattering processes
\cite{Abe97b} or the existence of new, low-energy contact interactions
\cite{Eic83}. The
latter could arise, for example, from the interchange of fermion
constituents at very short distances \cite{Lan92}.  A recent analysis of
$p{\bar p}\to \ell^+\ell^-$ data by the CDF collaboration  limits the
size of a lepton or quark to be $R< 5.6\times 10^{-4}$ f when $R$ is
determined from the assumed presence of a form factor at the
fermion-boson vertex \cite{Abe97b}. More stringent limits on the distance scale
associated with compositeness are obtained from the assumption of new
contact interations governed by a coupling of strength $g^2=4\pi$.
 Collider experiments yield $R\sim 1/\Lambda < 6
\times 10^{-5}$ f, where $\Lambda$ is the mass scale associated with new
dimension six lepton-quark operators \cite{Abe97b}.

It is conventional to write the lowest dimenion contact interactions as
\begin{equation}
{\cal L}_\sst{COMP} = 4\pi\sum_{ij}{\eta_{ij}\over\Lambda_{ij}^2}
{\bar e}_i \Gamma e_i {\bar q}_j\Gamma q_j
\ \ \ ,
\end{equation}
where $\Gamma$ is any one of the Dirac matrices and $i,j$ denote the
appropriate fermion chiralities ({\em e.g.}, ${\bar e}_L e_R {\bar q}_L
q_R$ or ${\bar e}_L\gamma_\mu e_L {\bar q}_R \gamma^\mu q_R$ {\em etc.}).
For simplicity, we restrict our attention to $\Gamma=\gamma^\mu$.
The quantities $\eta_{ij}$ take on the values $\pm 1, 0$ depending on
one's model assumptions. In terms of the PV interaction of Eq.
(\ref{eq:lnew}), the
contribution from ${\cal L}_\sst{COMP}$ is
\begin{equation}
\pi {\bar e}\gamma_\mu \gamma_5 e
\sum_{q}\left[{\eta_{RR}\over\Lambda_{RR}^2}-{\eta_{LL}\over\Lambda_{LL}^2}
+{\eta_{RL}\over\Lambda_{RL}^2}-{\eta_{LR}\over\Lambda_{LR}^2}\right]
{\bar q}\gamma^\mu q\ \ \ .
\end{equation}
Writing this interaction in terms of a common mass scale $\Lambda$ yields
\begin{equation}
{\pi\over\Lambda^2}{\bar e}\gamma_\mu\gamma_5 e\sum_q\left[
{\tilde\eta}_{RR}-{\tilde\eta}_{LL}+{\tilde\eta}_{RL}-{\tilde\eta}_{LR}
\right] \ \ \  ,
\end{equation}
where
\begin{equation}
{\tilde\eta}_{ij}=\eta_{ij}\left({\Lambda\over\Lambda_{ij}}\right)^2
\ \ \ .
\end{equation}
The correspondence with ${\cal L}_\sst{NEW}^{PV}$ is given by
\begin{eqnarray}
\kappa^2&=&1/4 \\
\hvq&=&{\tilde\eta}_{RR}-{\tilde\eta}_{LL}+{\tilde\eta}_{RL}-
{\tilde\eta}_{LR}
\end{eqnarray}

On the most general grounds, one has no strong argument for any of the
$\hvq$ to vanish. Consequently, low energy observables will generate lower
bounds on $\Lambda$. To compare with the recent CDF limits, we consider the
case of ${\tilde\eta}_{LL}=\pm 1$ and ${\tilde\eta}_{RR}={\tilde\eta}_{RL}=
{\tilde\eta}_{LR}=0$. In this case, the cesium APV results yield
\begin{equation}
\label{eq:apvcomp}
\Lambda_{LL}\geq 17.3 \ \ {\hbox{TeV}}
\end{equation}
assuming $\hvu=\hvd=-{\tilde\eta}_{LL}$.  Regarding other low-energy PV
observables, we note that the general comparisons made in Section III apply
here. Hence, a 10\% measurement of $\delta_P$ with PV $ep$ scattering
would yield comparable bounds, while a measurement of the isotopte ratio
${\cal R}_1$ with 0.5\% precision would be required to obtain comparable
limits.
Were the cesium APV theory error reduced to the level of the present
experimental
error, or were a 2-3\% determination of $\delta_P$ achieved, the lower
limit (\ref{eq:apvcomp})
would double.

Specific sensitivities from present and prospective measurements are given
in Table IV below:

$$\hbox{\vbox{\offinterlineskip
\def\strut{\hbox{\vrule height 15pt depth 10pt width 0pt}}
\hrule
\halign{
\strut\vrule#\tabskip 0.2cm&
\hfil$#$\hfil&
\vrule#&
\hfil$#$\hfil&
\vrule#&
\hfil$#$\hfil&
\vrule#\tabskip 0.0in\cr
& \multispan5{\hfil\bf TABLE IV\hfil} & \cr\noalign{\hrule}
& \hbox{Observable} && \hbox{Precision} && \Lambda_{LL}\ ({\hbox{TeV}}) &
\cr\noalign{\hrule}
& \qpv({\hbox{Cs}}) && 1.3\% && 17.3 & \cr
& && 0.35\% && 33.3 &\cr
& {\cal R}_1 &&  0.3\% && 21.6 &\cr
& && 0.1\% && 37.4 &\cr
& \qpv(^1{\hbox{H}})/Q_\sst{EM}(^1{\hbox{H}}) && 10\% && 17.5&\cr
& && 3\% && 31.8 &\cr
& \qpv(0^+,0)/Q_\sst{EM}(0^+,0) && 1\% && 21.6 &\cr
& \qpv(e)/Q_\sst{EM}(e) && 7\% && 17.1^{\hbox{a}} &\cr
&{\tilde a}_1 && 1\% && 2.6&\cr
\noalign{\hrule}}}}$$
{\noindent\narrower {\bf Table IV.} \quad Present and prosepctive limits on
compositeness scale for the \lq\lq LL" scenario. (a) M\"oller limits refer
to new $ee$ compositeness interactions, while other enteries refer to
$eq$ interactions.
\smallskip}

As with other new physics scenarios, the present and prospective low-energy
limits
on compositeness are competitive with those presently obtainable from collider
experiments as well as those expected in the future. The CDF collaboration has
obtained lower bounds on $\Lambda_{LL}(eq)$ of 2.5 (3.7) TeV for
${\tilde\eta}_{LL}=+1\
(-1)$ \cite{Abe97b}. One expects to improve these bounds to 6.5 (10) TeV
with the completion
of Run II  and 14 (20) TeV with TeV33 \cite{Bod96}. It is conceivable that
future improvements
in determinations of $\qpv$ with APV or scattering will yield stronger
bounds that those
expected from colliders. In the case of $\Lambda_{LL}(ee)$, $Z$-pole
observables imply
lower bounds of 2.4 (2.2) TeV for ${\tilde\eta}_{LL}=+1\ (-1)$ \cite{Ack97}.
The prospective M\"oller PV lower bounds exceed the LEP limits considerably.

The strength of these low-energy PV bounds has inspired various proposals
for evading
them. These scenarios include requiring ${\cal L}_\sst{COMP}$ to be parity
invariant \cite{Buc97}
(${\tilde\eta}_{RR}={\tilde\eta}_{LL}$,
${\tilde\eta}_{RL}={\tilde\eta}_{LR}$) or to
satisfy SU(12) symmetry \cite{Nel97} (in effect,
${\tilde\eta}_{iL}=-{\tilde\eta}_{iR}$, that is,
the new quark currents are purely axial vector).

\section{Theoretical Uncertainties}
\label{sec:theo}

The PV M\"oller asymmetry is the theoretically cleanest low-energy PV new
physics probe.
The dominant theoretical uncertainties are associated with hadronic
contributions to the
$Z-\gamma$ mixing tensor, and they do not appear to be problematic for the
extraction of
new physics limits \cite{Cza96}.
The attainment of stringent limits on new physics scenarios from low-energy
semi-leptonic
PV observables, however, requires that conventional many-body physics of
atoms and hadrons be
sufficiently well understood. At present, the dominant uncertainty in
$\qpv({\hbox{Cs}})$ is theoretical. A significant improvement in the
precision with which this
quantity is known requires considerable progress in atomic theory. The issues
involved in reducing the atomic theory uncertainty are discussed elsewhere
\cite{Woo97,Dzu89,Blu92}. In this section, we discuss the many-body
uncertainties associated
with the other semi-leptonic observables discussed above.

\medskip
\noindent{\bf A. Isotope ratios}
\medskip

It was pointed out in Refs. \cite{For90,Pol92} that the isotope ratios
${\cal R}_i$ display an
enhanced sensitivity to the neutron distribution $\rho_n(r)$ within atomic
nuclei, and that
uncertainties in $\rho_n(r)$ could hamper the extraction of new physics
limits from the ${\cal
R}_i$. In Ref. \cite{Pol92}, only ${\cal R}_2$ was considered, and only the
implications of
$\rho_n(r)$ uncertainties for the determination of $\sstw$ were discussed.
For completeness, we
consider also
${\cal R}_1$ -- which displays a greater new physics sensitivity than
${\cal R}_2$ -- and
quantify the implications of $\rho_n(r)$ uncertainties for the extraction
of new physics limits.

In general, one may express the weak charge as
\begin{equation}
\qpv= Z \qwp q_p + N\qwn q_n \ \ \ ,
\end{equation}
where
\begin{eqnarray}
q_p&=& (1/{\cal N})\int\ d^3x \bra{P}{\hat\psi}_e^{\dag}({\vec
x})\gamma_5{\hat\psi}_e({\vec x})
\ket{S}\rho_p({\vec x})\\
\label{eq:apvqn}
q_n&=& (1/{\cal N})\int\ d^3x \bra{P}{\hat\psi}_e^{\dag}({\vec
x})\gamma_5{\hat\psi}_e({\vec x})\ket{S}
\rho_n({\vec x})
\end{eqnarray}
where ${\hat\psi}_e({\vec x})$ is the electron field operator, $\ket{S}$
and $\ket{P}$ are
atomic $S_{1/2}$ and $P_{1/2}$ states, and  ${\cal N}$ is the value of the
electron matrix
element at the origin. The latter matrix element may be written as
\begin{equation}
\bra{P}{\hat\psi}_e^{\dag}({\vec x})\gamma_5{\hat\psi}({\vec
x})\ket{S}={\cal N} f(x) \ \ \ ,
\end{equation}
where $f(0)=1$. The effect of uncertainties in $\rho_p({\vec x})$ -- which
are smaller than those
in $\rho_n({\vec x})$ -- are suppressed in $\qpv$ since $q_p$ is multiplied
by the small
number $\qwp$. Consequently, we consider only $q_n$.

To obtain general features, we follow Refs. \cite{For90,Pol92} and consider
a simple model in
which the nucleus is treated as a sphere of uniform proton and neutron
number densities out to
radii $R_P$ and $R_N$, respectively.  In this case,  one obtains\cite{Pol92}
\begin{equation}
q_n=1-(Z\alpha)^2 f_2^N+\cdots\ \ \ ,
\end{equation}
where
\begin{eqnarray}
\label{eq:f2n}
f_2^N&=& {3\over 10}x_N^2 -{3\over 70} x_N^4 +{1\over 450} x_N^6\\
x_N&=& R_N/R_P \ \ \ .
\end{eqnarray}
Letting $\delta^n_N$ and $\delta^n_i$ denote the $\rho_n({\vec x})$
corrections to $\qpv(N)$ and
${\cal R}_i$, respectively, we obtain
\begin{eqnarray}
\delta^n_N&\approx& -(Z\alpha)^2 f_2^N(x_N)\\
\delta^n_1&\approx& -(Z\alpha)^2 (N'/\Delta N) f_2^{N\prime}(x_N)\Delta x_N\\
\delta^n_2&\approx& -(Z\alpha)^2 f_2^{N\prime}(x_N) \Delta x_N\ \ \ ,
\end{eqnarray}
where $\Delta x_N=(R_{N'}-R_N)/R_P$. Uncertainties in $\qpv$ and ${\cal
R}_i$ arise from
{\em uncertainties} in these quantities:
\begin{eqnarray}
\label{eq:uncer1}
\delta (\delta_N^n) &\approx & -(Z\alpha)^2 f_2^{N\prime}(x_N) \delta x_N\\
\label{eq:uncer2}
\delta (\delta_1^n)&\approx & -(Z\alpha)^2(N'/\Delta
N)\left[f_2^{N\prime}(x_N)\delta(\Delta x_N)
  +(\Delta x_N)f_2^{N\prime\prime}(x_N)\delta x_N\right]\\
\label{eq:uncer3}
\delta (\delta_2^n)&\approx &
-(Z\alpha)^2\left[f_2^{N\prime}(x_N)\delta(\Delta x_N)
  +(\Delta x_N)f_2^{N\prime\prime}(x_N)\delta x_N\right] \ \ \ ,
\end{eqnarray}
where $\delta x_N$ is the uncertainty in $x_N$ {\em etc.}

From the standpoint of extracting new physics limits, the impact of neutron
distribution
uncertainties is characterized by the ratio of the $\delta (\delta^n_k)$ to
the new
physics corrections $\delta_k$ ($k=N, 1, 2$). The smaller the size of this
ratio, the
less problematic neutron distribution uncertainties become. In the case of
the isotope
ratios, we observe that
\begin{equation}
\label{eq:del1eqdel2}
\delta (\delta^n_1)/\delta_1\approx -\left({N'\over\Delta N}\right){(Z\alpha)^2\over\zeta}
{\left[f_2^{N\prime}(x_N)\delta(\Delta x_N)
  +(\Delta x_N)f_2^{N\prime\prime}(x_N)\delta x_N\right]\over2\hvu+\hvd}
\approx\delta (\delta^n_2)/\delta_2 \ \ \ .
\end{equation}
In short, the relative size of the corrections induced by new physics and
neutron distribution
uncertainties is essentially the same, whether one employs ${\cal R}_1$ or
${\cal R}_2$. Although
${\cal R}_1$ is more sensitive to new physics by $N'/\Delta N$ as compared
to ${\cal R}_2$, it
is also more sensitive to $\rho_n({\vec x})$ uncertainties by the same factor.

To set the scale of $\rho_n({\vec x})$ uncertainties, we set $x_N\approx 1$
in Eqs.
(\ref{eq:f2n}-\ref{eq:uncer3}):
\begin{eqnarray}
\delta (\delta_N^n)&\approx& -(3/7) (Z\alpha)^2 \ \delta x_N\\
\delta (\delta_1^n)&\approx& -(N'/\Delta N) (Z\alpha)^2\
\left[(3/7)\delta(\Delta x_N)+
(1/8)\Delta x_N
\ \delta x_N\right]\\
\delta (\delta_2^n)&\approx& - (Z\alpha)^2\ \left[(3/7)\delta(\Delta x_N)+
(1/8)\Delta x_N
\ \delta x_N\right]\ \ \ .
\end{eqnarray}
In general, one has $\Delta x_N\delta x_N << \delta(\Delta x_N)$
\cite{Pol92}. Consequently,
we keep only the terms associated with the uncertainty in the isotope
shift, $\delta(\Delta x_N)$.

We specify these expressions for the case of Cs, Yb, Ba, and Pb. Although
no studies of
cesium isotope ratios are planned at present, we include it in order to
make a direct comparison
between the single isotope and isotope ratios for this atom. The Yb and Ba
isotopes are under
study by the Berkeley and Seattle groups, respectively. We also include
lead since it is
one of the best understood heavy nuclei, both experimentally and
theoretically. The neutron
distribution uncertainties are shown for $\qpv(^{133}{\hbox{Cs}})$ and for
${\cal R}_1$ for Cs,
Yb, Ba, and Pb.  In light of Eq. (\ref{eq:del1eqdel2}), it is sufficient to
consider only ${\cal
R}_1$. The fourth column of Table V gives the requirement on neutron
distribution uncertainties
for a given uncertainty in the corresponding APV observable. For
$\qpv({\hbox{Cs}})$, we require
$\delta (\delta_N^n)$ to be smaller than the present experimental
uncertainty. For the
isotope ratios, the requirement is $\delta (\delta_1^n) \leq 0.1\% $. In
either case, the
requirement must be met if the present cesium APV new physics reach is to
be doubled. In the
final column, we list published theoretical esimates of the corresponding
neutron distribution
uncertainty. The range in the case of ${\cal R}_1({\hbox{Cs}})$ corresponds
to using the
nominal error of Ref. \cite{Che93} (larger value) and the spread between
two models used in
the calculation (lower value).

$$\hbox{\vbox{\offinterlineskip
\def\strut{\hbox{\vrule height 15pt depth 10pt width 0pt}}
\hrule
\halign{
\strut\vrule#\tabskip 0.2cm&
\hfil$#$\hfil&
\vrule#&
\hfil$#$\hfil&
\vrule#&
\hfil$#$\hfil&
\vrule#&
\hfil$#$\hfil&
\vrule#&
\hfil$#$\hfil&
\vrule#\tabskip 0.0in\cr
& \multispan9{\hfil\bf TABLE V\hfil} & \cr\noalign{\hrule}
& \hbox{Observable} && \hbox{Precision} && \Delta N && \hbox{Requirement} &&
\hbox{Theory}
& \cr\noalign{\hrule}
& \qpv({\hbox{Cs}}) && 0.35\% && 0 && \delta x_N\leq 0.05 && \delta x_N\leq
0.02^a& \cr
& {\cal R}_1({\hbox{Cs}}) && 0.1\%  && 14 && \delta (\Delta x_N)\leq 0.0024
&&(\Delta x_N)\leq
0.0033\to 0.0043^b  &\cr
 & {\cal R}_1({\hbox{Ba}}) && 0.1\%  && 14 && \delta (\Delta x_N)\leq
0.0023 && (\Delta x_N)\leq
0.0038^c  &\cr
& {\cal R}_1({\hbox{Yb}}) && 0.1\%  && 6 && \delta (\Delta x_N)\leq 0.0005
&& -- &\cr
& {\cal R}_1({\hbox{Pb}}) && 0.1\%  && 6 && \delta (\Delta x_N)\leq 0.0003
&& (\Delta x_N)\leq
0.005^d &\cr
\noalign{\hrule}}}}$$
{\noindent\narrower {\bf Table V.} \quad Neutron distribution uncertainties
in atomic
parity violation. First line gives results for $^{133}$Cs and following
four give results
for isotope ratios. The isotope spread $\Delta N$ is taken from Ref.
\cite{Vog94} for
Cs and Ba, from Ref. \cite{Bud98} for Yb, and from Ref. \cite{Pol92} for
Pb. Fourth
column gives required precision in neutron radius and isotope shift in
order to keep
neutron distribution uncertainty below the level quoted in column three.
Fifth column
gives theoretical estimates of neutron distribution uncertainties: (a,b)
Ref. \cite{Che93},
(c) Ref. \cite{Vog94}, (d) ref. \cite{Pol92}.
\smallskip}

At present, there exist no reliable experimental determinations of $x_N$ or
$\Delta x_N$, so that the interpretation of APV observables must rely on
nuclear
theory\footnote{Data from proton-nucleus and pion-nucleus exist for some
cases, but the
theretical uncertainties are large. See, {\em e.g.}, Ref. \cite{Pol92}.}.
It is conceivable that
the theory uncertainty in $x_N$ is 5\% or better \cite{Pol92,Che93}. The
estimate of Ref.
\cite{Che93} places this uncertainty closer to 2\%. Consequently, one could
argue that even
if the atomic theory error in
$\qpv({\hbox{Cs}})$ were reduced to the present experimental error, neutron
distribution
uncertainties should not complicate the extraction of new physics
constraints. The situation
regarding isotope shifts is more debatable.

Explicit studies of isotope shift uncertainties associated with $\rho_n(r)$
have been reported
in Refs. \cite{For90,Pol92,Che93,Vog94}. The authors of Ref. \cite{Pol92}
considered isotopes of
lead  using a variety of nuclear models and find a model spread of
$\delta(\Delta x_N)\approx
0.005$, which corresponds to a 100\% uncertainty in the model average for
$\Delta x_N$. These
authors note that the models used successfully predict the charge radii of
even-even nuclei not
used to fit the model parameters. The model spread is a factor of ten
larger than would be needed
to keep the uncertainty in ${\cal R}_1({\hbox{Pb}})$ below 0.1\%. Although
the isotopes of lead
are not presently under serious consideration for isotope ratio
measurements, the scale of the
model uncertainties for this well-understood set of isotopes is striking.

The authors of Ref. \cite{Che93} employed two different Skyrme
fits to compute $\Delta x_N$ for Cs and Ba and quote an uncertainty in
$\Delta x_N$ of
roughly 13\% for the two series of isotopes (in the case of cesium, the
difference in
$\Delta x_N$ between the two Skyrme fits is somewhat smaller than the
quoted uncertainty). To our
knowledge, there exist no published analyses of the
$\rho_n$ uncertainties for Yb. From the studies of Pb, Cs, and Ba, we infer
that $\rho_n$
uncertainties are presently larger than required for isotope ratio
measurements to compete with those on a single isotope for yielding new
physics limits.

Obtaining a sufficiently reliable computation of $\Delta x_N$ remains an open
problem for nuclear theory. It is argued in Ref. \cite{Pol92}, for example,
that model
calculations contain a hidden uncertainty associated with the isovector
surface term in the
nuclear energy functional. Changes in the coefficient of this term may
signficantly affect a model
calculation of $\Delta x_N$ without affecting results for other
observables. The authors
of Ref. \cite{Che93}, on the other hand, considered this issue for cesium
using a Skyrme
interaction with two different parameter sets. For this interaction,
changes in the isovector
surface term larger enough to appreciably alter $\Delta x_N$ also produce
unacceptably large
changes in binding energies. Whether the Skyrme results generalize to other
interactions remains
to be seen.

Given the present theoretical situation, a model-independent determination
of $\rho_n({\vec x})$
is desirable. To that end, PVES may prove useful \cite{Don89,Mus94}.
Specifically, we
consider a
$(J^\pi, T)=(0^+,0)$ nucleus, such as $^{138}_{56}$Ba, noting that the
isotopes of barium
are under consideration for future APV isotope ratio measurements. As shown
in Refs.
\cite{Don89,Mus94}, the PV asymmetry for $(0^+,0)$ nuclei may be written as
\begin{equation}
-\left[{4\sqrt{2}\pi\alpha\over G_F|q^2|}\right]\alr = \qwp+\qwn{\int\
d^3x\ j_0(qx)
\rho_n({\vec x}) \over \int\ d^3x\ j_0(qx) \rho_p({\vec x})}\ \ \ .
\end{equation}
Since $|\qwp/\qwn| << 1$, and since $\rho_p({\vec x})$ is generally well
determined from
parity conserving electron scattering, $\alr$ is essentially a direct
\lq\lq meter" of the
Fourier transform of $\rho_n({\vec x})$. At low momentum-transfer
($qR_{N,P}<< 1$) this
expression simplifies:
\begin{equation}
-\left[{4\sqrt{2}\pi\alpha\over G_F|q^2|}\right]\alr\approx {N\over
Z}\left[1+{q^2\over 6}\left(
{R_P^2\over Z}-{R_N^2\over N}\right)\right]
\end{equation}
so that a determination of $R_N$ is, in principle, attainable from
$\alr$\footnote{In a realistic
analysis of $\alr$ for heavy nuclei, the effects of electron wave
distortion must be included
in the analysis of $\alr$. For a recent distorted wave calculation, see
Ref. \cite{Hor98}.}.

In a realistic experiment PVES experiment, one does not have $qR_{N,P} <<
1$; larger values of
$q$ are needed to obtain the requisite precision for reasonable running times
\cite{Don89,Mus94}. In Ref. \cite{Don89}, it was shown that a 1\%
determination of
$\rho_n({\vec x})$ for $^{208}$Pb is experimentally feasible for $q\sim
0.5\ {\hbox{fm}}^{-1}$
with reasonable running times. An experiment with barium is particularly
attractive. If the
barium isotopes are used in future APV measurements as anticipated by the
Seattle group,
then a determination of $\rho_n({\vec x})$ for even one isotope could
reduce the degree of
theoretical uncertainty for neutron distributions along the barium isotope
chain. Moreover, the
first excited state of
$^{138}$Ba occurs at 1.44 MeV. The energy resolution therefore required to
guarantee elastic
scattering from this nucleus is well within the capabilities of the
Jefferson Lab.

The foregoing discussion illustrates general features of, and presents order of
magnitude estimates for, neutron distribution effects in APV. A more
complete analysis
of $q_n$ using realistic atomic wavefunctions will be required to translate
PVES information
on $\rho_n(q)$ into useful input for APV calculations. Indeed, the function
$f(x)$ which
weights $\rho_n({\vec x})$ in Eq. (\ref{eq:apvqn}) is not the same as the
Bessel function
$j_0(qx)$ which weights $\rho_n$ in the asymmetry. Evidently, a
determination of $\rho_n(q)$ over
some range in
$q$ will be required. Assuming $\rho_n({\vec x})$ can be sufficiently well
determined for a
single isotope, it remains to be seen how tightly such a determination
constraint nuclear
theory calculations of $\rho_n({\vec x})$ along the isotope chain or
elsewhere in the periodic
table. A detailed treatment of these issues lies beyond the scope of the
present study.

\medskip
\noindent{\bf B. Hadronic Form Factors}
\medskip

From the form of Eq. (\ref{eq:alr}), it is clear that a precise
determination of $\qpv$ from
$\alr$ requires sufficiently precise knowledge of the form factor term,
$F(q)$. This term is
presently under study at a variety of accelerators, with the hope of
extracting information on
the strange quark matrix element $\bra{N(p')}{\bar s}\gamma_\mu
s\ket{N(p)}$. The latter is
parameterized by two form factors, $\GES$ and $\GEp$. The other form
factors which enter $F(q)$
are known with much greater certainty than are the strange quark form
factors. A separation of
$\qpv$ from $F(q)$ requires at least one forward angle measurement
\cite{Mus92}. The kinematics
must be chosen so as to minimize the importance of $F(q)$ relative to
$\qpv$ while keeping the
statistical uncertainty in the asymmetry sufficiently small. These
competing kinematic
requirements -- along with the desired uncertainty in $\qpv$ -- dictate the
maximum uncertainty
in $F(q)$ which can be tolerated. Since $\alr(^1H)$ generally manifests the
greatest sensitivity
to new physics, we illustrate the form factor considerations for PV $ep$
scattering.

Since $Q_\sst{EM}^p=1$, the $ep$ asymmetry has the form
\begin{equation}
\alr=a_0\tau\ \left[\qwp+F^p(q)\right]\ \ \ ,
\end{equation}
where $a_0\approx -3.1\times 10^{-4}$ and $\tau=|q^2|/4\mns$. The form
factor contribution
is given at tree level in the Standard Model by \cite{Mus92,Mus94}
\begin{equation}
F^p(\tau)=-\left[\GEp(\GEn+\GES)+\tau\GMp(\GMn+\GMS)\right]/\left[(\GEp)^2+\tau(
\GMp)^2\right]
\ \ \ ,
\end{equation}
where $G^{p,n}_\sst{E,M}$ denote the proton or neutron Sachs electric or
magnetic form factors.
Since $Q_\sst{EM}^n=0$ and since the proton carries no net strangeness,
both $\GEn$ and
$\GES$ must vanish at $\tau=0$. Consequently, we may write $F^p(\tau)$ as
\begin{equation}
F^p(q)=\tau B(\tau)\ \ \ .
\end{equation}

For purposes of this discussion, it is useful to write
\begin{equation}
B(\tau)=B_0(\tau)+\delta B(\tau)\ \ \ ,
\end{equation}
where $B$ gives the contribution from $\GEp$, $\GEn$, $\GMp$, and $\GMn$
while $\delta B$
contains the contributions from $\GES$ and $\GMS$ as well as from
$\tau$-dependent higher-order
electroweak corrections, as discussed in the next section. As noted in Ref.
\cite{Mus92},
any determination of $\qwp$ must be made at such low-$\tau$ that only
$B(\tau=0)$ enters
the analysis. The experimental problem is to measure $B(\tau)$ with
sufficient precision over
a sufficient range of $\tau$ such that $\tau\delta B(0)$ smaller than the
desired uncertainty
in $\qwp$ in a low-$\tau$ measurement.

In the first 12 lines of Table VI, we summarize the conditions for several
prospective
determinations of $\qwp$ with a measurement of $\alr(^1{\hbox{H}})$. The
last three lines
summarize existing or planned forward angle determinations of $B(\tau)$.
For both sets of
measurements, the third column gives the statistical uncertainty in the
asymmetry, assuming
a solid angle of 10 msr, a luminosity ${\cal L}= 5\times 10^{38}$
cm$^{-2}$s$^{-1}$ and
100\% beam polarization for various running times and
kinematics\cite{Mus94}. The fourth column
gives the corresponding experimental uncertainty in $\qwp$. The
requirements to keep the error
in $\qwp$ from $B(\tau)$ smaller than the statistical uncertainty are given
in the fifth column.
For the second set of measurements (final three rows), only the Standard
Model uncertainty in
$\qwp$ is listed. The dominant uncertainty arises from hadronic loops
appearing in the
$Z-\gamma$ mixing tensor \cite{Cza96}. The uncertainty associated with the
experimental
value of $\sstw$ is about a factor of three smaller than the hadronic
uncertainty \cite{PDG98}.
We note that measurements at $\theta=6^\circ$ would require the development
of new beam optics
for the CEBAF detectors; such developments appear technically
feasible\cite{PV97,Car97}. The
choices for $\tau$ in the first 12 lines correspond roughly to CEBAF beam
energies.

$$\hbox{\vbox{\offinterlineskip
\def\strut{\hbox{\vrule height 15pt depth 10pt width 0pt}}
\hrule
\halign{
\strut\vrule#\tabskip 0.2cm&
\hfil$#$\hfil&
\vrule#&
\hfil$#$\hfil&
\vrule#&
\hfil$#$\hfil&
\vrule#&
\hfil$#$\hfil&
\vrule#&
\hfil$#$\hfil&
\vrule#\tabskip 0.0in\cr
& \multispan9{\hfil\bf TABLE VI\hfil} & \cr\noalign{\hrule}
& \theta && \tau && \delta\alr/\alr && \delta\qwp/\qwp && \delta B/B_0  & \cr
\noalign{\hrule}
& 12.3^\circ && 0.018 && 0.063^a && 0.112 && 0.145 &\cr
& && && 0.045^b && 0.08 && 0.103 &\cr
& 6^\circ && 0.004 && 0.087^a && 0.103 && 0.59 &\cr
& && && 0.062^b && 0.073 && 0.42 &\cr
& 6^\circ && 0.008 && 0.059^a && 0.08 && 0.227 &\cr
& && && 0.041^b && 0.056 && 0.161 &\cr
& 6^\circ && 0.012 && 0.043^a && 0.065 && 0.123 &\cr
& && && 0.033^b && 0.046 && 0.087 &\cr
& 6^\circ && 0.018 && 0.032^a && 0.057 && 0.074 &\cr
& && && 0.023^b && 0.04 && 0.052 &\cr
& 6^\circ && 0.024 && 0.025^a && 0.05 && 0.05 &\cr
& && && 0.018^b && 0.035 && 0.035 &\cr
\noalign{\hrule}
&12.3^\circ && 0.14 && 0.16^c && 0.032^e && 0.196 &\cr
&12.3^\circ  && 0.14 && 0.05^c && 0.032^e && 0.065 &\cr
& 35^\circ && 0.07 && 0.04^d && 0.032^e  && 0.057 &\cr
\noalign{\hrule}}}}$$
{\noindent\narrower {\bf Table VI.} \quad Conditions for new physics search
with PV elastic $ep$ scattering. First 12 lines give conditions for
determination
of $\qwp$ with the precision listed in column four asssuming 10 msr solid
angle detector,
${\cal L}= 5\times 10^{38}$ cm$^{-2}$s$^{-1}$, 100\% beam polarization and
(a) 1000 hours
of running time (b) 2000 hours of running time. The corresponding
statistical uncertainty
in the asymmetry is listed in column three, while the required precision in
$B(\tau)$ is given
in column five. Final three lines give present present and prospective
determinations of
$B(\tau)$ at Jefferson Laboratory (c) (See Ref. \cite{Ani98,Fin91}) and
Mainz (d) (see Ref.
\cite{Har91}). (e) The uncertainty in $\qwp$ in the last three lines is
computed using
the hadronic uncertainty of Ref. \cite{Cza96}.
\smallskip}

The results entries in Table VI illustrate the trade-offs between
kinematics, desired
precision in $\qwp$, and required precision in $B(\tau)$. For a given
scattering angle
$\theta$, increasing $\tau$ decreases the statistical uncertainty in $\alr$
but increases
the contribution from $\tau B(\tau)$. The latter increase has two effects.
First, it
reduces the relative contribution of $\qwp$, making it more difficult to
match the
fractional uncertainty in $\alr$ with the desired uncertainty in $\qwp$.
Second, it imposes
more stringent requirements on knowledge of $B(\tau)$. Consequently, it may be
desireable to go to slightly longer running times and lower $\tau$.
Comparing the two possible
measurements at $\theta=6^\circ$, for example, we see that a 1000 hour
measurement at $\tau=0.024$
yields a 2.5\% statistical uncertainty in $\alr$ but only a 5\% uncertainty
in $\qwp$. Moreover,
the required precision on $B$ is slightly more stringent than will be
obtained with any of the
current PVES measurements (last three lines). However, a 2000 hour
$\theta=6^\circ$ experiment
at $\tau=0.018$ yields a 4\% determination of $\qwp$ for a 2.3\%
measurement of $\alr$ while
imposing similar requirements on $\delta B$. An even more precise
determination of $\qwp$ would
require reduction in the hadronic uncertainty entering the Standard Model
radiative corrections.

We emphasize that the entries in Table VI are intended as illustrative
benchmarks. The optimal
kinematics for a precise determination of $\qwp$ require a detailed analysis of
acutal experimental conditions at different laboratories. We also emphasize
that the
measured uncertainty in $B$ at higher $\tau$ (last three lines of Table VI)
does not
necessarily translate into the same uncertainty at the lower $\tau$ needed
for new physics
searches. For example, the strange quark form factors may not scale with
$\tau$ in the same
way as the nucleon EM form factors. Hence,
it is likely that measurements of $B(\tau)$ over a range of kinematics will be
needed to sufficiently constrain its value at the photon point (see, {\em
e.g.}, Refs.
\cite{Mus92,Mus94}. A detailed analysis of this issue would constitute a
critical component
of an experimental proposal.

\medskip
\noindent{\bf C. Dispersion Corrections}
\medskip

The foregoing discussion has implicitly relied upon a first Born
approximation of the
electroweak amplitudes contributing to low-energy PV. A realistic analysis
of precision
observables must take into account contributions beyond the first Born
amplitude. In the case
of electron scattering, these contributions are generally divided into two
classes: Coulomb
distortion of plane wave electron wavefunctions and dispersion corrections.
The former can
be treated accurately for electron scattering using distorted wave methods.
Results of such a
treatment are reported in Ref. \cite{Hor98}. The dispersion correction,
however, has proven less
tractable.

The leading dispersion correction (DC) arises from diagrams of Fig. 2,
where the
intermediate state nucleus or hadron lives in any one of its excited
states. More generally,
box diagrams like those of Fig. 2 can be treated exactly for scattering of
electrons
from point like hadrons. When at least one of the exchanged bosons is a
photon, the amplitude
is prone to infrared enhancements.  For elastic PV scattering of an
electron from a point-like
proton, for example, the $Z-\gamma$  amplitude contains infrared
enhancement factors such as
$\ln|s|/\mzs$, where $s$ is the $ep$ c.m. energy \cite{Mus90}.  Such
factors can enhance
the scale of the amplitude by as much
as an order of magnitude over the nominal ${\cal O}(\alpha)$ scale.
Consequently,  one might
expect box graph amplitudes which depend on details of hadronic or nuclear
structure to be a
potential source of theoretical error in the analysis of precision
electroweak observables.

Data on the electromagnetic  ($\gamma\gamma$) dispersion correction for
$ep$ scattering
is in general agreement with the scale predicted by  theoretical calculations.
The situation regarding electron scattering from nuclei, however, is less
satisfying.
Recent data $^{12}$C$(e,e')$ taken at MIT-Bates and NIHKEF disagree
dramatically
with nearly all published calculations (for a more detailed discussion and
references,
see Ref. \cite{Mus93b}). An experimental determination of any electroweak
DC ($V=\gamma$,
$V'=W^\pm,Z^0$) is unlikely, and reliance on theory to compute this
correction is unavoidable.
As we show below, the corresponding theoretical uncertainty is far less
problematic for
a determination of $\qpv$ from PVES than for the extraction of information
on the strange
quark form factors.

To this end, it is convenient to write the $(V,V')$ DC as a correction
$R_{VV'}$ to the
tree level EM and PV neutral current ampltitudes \cite{Mus93b}:
\begin{eqnarray}
\label{eq:dcampl}
M_\sst{EM}&=&M_\sst{EM}^\sst{TREE}[1+R_{\gamma\gamma}+\cdots]\\
M_\sst{NC}^\sst{PV}&=&M_\sst{NC}^\sst{PV,\ TREE}[1+R_{VV'}+\cdots]\ \ \ ,
\end{eqnarray}
where $\cdots$ denotes other higher order corrections to the tree level
amplitude. Because
$M_\sst{EM}^\sst{TREE}\propto 1/q^2$ while the $\gamma\gamma$ amplitude
contains no pole
at $q^2=0$, $R_{\gamma\gamma}$ has the general structure
\begin{equation}
\label{eq:rtildef}
R_{\gamma\gamma}(q^2)=q^2\ {\tilde R}_{\gamma\gamma}(q^2)
\end{equation}
where ${\tilde R}_{\gamma\gamma}(q^2)$ describes the $q^2$ dependence of
the $\gamma\gamma$
amplitude and ${\tilde R}_{\gamma\gamma}(0)$ is finite.
Since the tree level NC amplitude contains no pole at $q^2=0$, however, the
PV DC's do not vanish at $q^2=0$. Using Eqs.
(\ref{eq:dcampl}-\ref{eq:rtildef}) and expanding the
PV corrections in powers of
$q^2$ we obtain
\begin{equation}
\label{eq:alrdc}
{\alr\over a_0\tau}= \qpv\left[1+R_{WW}(0)+R_{ZZ}(0)+R_{Z\gamma}(0)\right]
+ {\tilde F}(q)
\end{equation}
where we replace the form factor $F(q)$ appearing in Eq. (\ref{eq:alr}) by
an effective
form factor ${\tilde F}(q)$:
\begin{equation}
{\tilde F}(q)=F(q)+q^2\left[
R^{\prime}_{WW}(0)+R^{\prime}_{ZZ}(0)+R^{\prime}_{Z\gamma}(0)-{\tilde
R}_{\gamma\gamma}(q^2)
+\cdots\right]\ \ \ ,
\end{equation}
with $F(q)$ containing the dependence on hadronic form factors as before.

From Eq. (\ref{eq:alrdc}) we observe that the entire $\gamma\gamma$ DC, as
well as the sub-leading
$q^2$-dependence of the $WW$, $ZZ$, and $Z\gamma$ DC's, contribute to
$\alr$ as part of
an effective form factor term, ${\tilde F}(q)$. Since $F(q)\sim q^2$ for
low-$|q^2|$ at
forward angles, the DC contributions entering Eq. (\ref{eq:alrdc}) will be
exerimentally
constrained along with $F(q)$ when the form factor term ${\tilde F}(q)$ is
kinematically
separated from the weak charge term. Consequently, an extraction of $\qpv$
from $\alr$ does not
require  theoretical computations of the $\gamma\gamma$ DC or of the sub-leading $q^2$-dependence
of the other DC's. A determination of the strange-quark for
m factors, however, does require
such theoretical input.

In order to constrain possible new physics contributions to $\qpv$, a
Standard Model theoretical
calculation of $R_{WW}(0)$, $R_{ZZ}(0)$, and $R_{Z\gamma}(0)$ is
necessary.The theoretical
uncertainty associated with $R_{WW}(0)$ and $R_{ZZ}(0)$ is small, since box
diagrams
involving the exchange are dominated by hadronic intermediate states having
momenta
$p\sim M_\sst{W}$.  These contributions can be reliably treated
perturbatively. The
$R_{Z\gamma}(0)$ correction, however, is infrared enhanced and displays a
greater sensitivity to
the low-lying part of the nuclear and hadronic spectrum. Fortunately, the
sum of diagrams 2a
and 2b conspire to suppress this contribution by $\gve=-1+4\sstw$. This
feature was first shown
in Ref. \cite{Mar83} for the case of APV. Here, we summarize the argument
as it applies to
scattering.

The dominant contributions to the loop integrals for diagrams 2a and 2b
arise when
external particle masses and momenta are neglected relative to the loop
momentum
$\ell_\mu$. In this case, the integrands from the two loop integrals sum to
give
\begin{eqnarray}
\label{eq:pvdc}
{\bar u}[\gamma_\alpha\not\ell\gamma_\beta(\gve+\gae\gamma_5)
  &-&\gamma_\beta(\gve+\gae\gamma_5) \not\ell\gamma_\alpha ] u\>
T^{\alpha\beta}(\ell) D(\ell^2) \\
&=& 2i\ \epsilon_{\alpha\lambda\beta\mu}\ell^\lambda{\bar
u}\gamma^\mu(\gve\gamma_5+\gae)u\>
T^{\alpha\beta}(\ell) D(\ell^2)\ \ \ ,\nonumber
\end{eqnarray}
where
\begin{equation}
T^{\alpha\beta}(\ell) = \int d^4x\ {\hbox{e}}^{i\ell\cdot x}\
\bra{0} T\left\{ J^\alpha_\sst{EM}(x) J^\beta_\sst{NC}(0)
\right\}\ket{0}\ \ \ ,
\end{equation}
$D(\ell^2)$ contains the electron and gauge boson propagators when external
momenta and
masses are neglected relative to $\ell_\mu$, and $J^\alpha_\sst{EM}$ and
$J^\beta_\sst{NC}$
are the hadronic electromagnetic and weak neutral currents, respectively.
The terms in Eq.
(\ref{eq:pvdc}) which transform like pseudoscalars are those containing the
EM current and either
(a) both the axial currents ${\bar u}\gamma^\mu\gamma_5 u$ and
$J_\sst{NC}^{\beta 5}$ or
(b) both the vector currents ${\bar u}\gamma^\mu u$ and $J_\sst{NC}^\beta$.
The former
has the coefficient $\gve=-1+4\sstw$ and the latter has a the coefficient
$\gae=1$. The
dependence of these terms on the spatial currents is given by ($\lambda=0$
in Eq. (\ref{eq:pvdc}))
\begin{eqnarray}
\gve\ {\hbox{term}}: &\sim& {\bar u}{\vec\gamma}\gamma_5 u\cdot\left({\vec
J}_\sst{EM}
	\times{\vec J}_\sst{NC}^5\right)\\
\gae\ {\hbox{term}}: &\sim& {\bar u}{\vec\gamma} u\cdot\left({\vec J}_\sst{EM}
	\times{\vec J}_\sst{NC}\right)\ \ \ .
\end{eqnarray}
The hadronic part of the $\gve$ term transforms as a polar vector, so that this
term contributes to the $A(e)\times V({\hbox{had}})$ amplitude. The
hadronic part
of the $\gae$ terms, on the other hand, transfors as an axial vector, yielding
a contribution to the $V(e)\times A({\hbox{had}})$ amplitude. Hence, only the
$\gve$ term contributes to $\qpv$ term in the asymmetry.

Since $\gve\sim -0.1$, the contribution $R_{Z\gamma}(0)$ in Eq.
(\ref{eq:alrdc}) is suppressed.
For scattering from $(0^+,0)$ nuclei, then, $R_{Z\gamma}(0)\sim {\cal
O}(\alpha/10)$, while
for PV $ep$ scattering, $R_{Z\gamma}\sim{\cal O}(\alpha)$. Since 1\% and
5-10\% determinations
of $\qpv(0^+,0)$ and $\qpv(p)$, respectively, are needed to constrain new
physics
scenarios, large theoretical uncertainties in $R_{Z\gamma}(0)$ should not
be problematic. A
similar statement applies to APV, for which contributions to $R_{Z\gamma}$
from excited
nuclear states have yet to be computed. Whether these contributions can be
reliably
computed at the 0.3\% level remains to be evaluated.

\section{Conclusions}
\label{sec:concl}

The prospects for future, precise measurements of low-energy PV observables
is promising.
In addition to the approved PV M\"oller experiment and planned APV isotope
measurements,
a precise measurement of $\alr$ for PV electron-proton or electron-nucleus
scattering
at Jefferson Laboratory appears feasible. Depending on the degree of
experimental and
theoretical precision realized in each case, future measurements could
improve upon the
present cesium APV new physics sensitivity by a factor of two. At the same
time, such studies
would complement future new physics searches at high-energy colliders.
Indeed, while
high-energy studies are particularly sensitive to the mass scale $\Lambda$
associated with
new interactions, low-energy PV probes the coupling-to-mass ratio,
$g/\Lambda$. For
new physics scenarios in which $g$ is fixed ({\em e.g.}, $LR$ symmetric
gauge theories or fermion
compositeness), even the present cesium APV bounds on $\Lambda$ exceed
those obtained from
the Tevatron or LEP2. Taken together, high-energy and low-energy PV
measurements provide a
powerful, combined probe of physics at the TeV scale.

As the discussion of Sections 3 and 4 illustrates, no single low-energy PV
process is equally
sensitive to every new physics scenario. For example, APV on a single
isotope is strongly
sensitive to new isoscalar interactions but much less transparent to new
isovector heavy physics.
Similarly, elastic PV $ep$ scattering constitutes the most sensitive probe
of new $e-q$ physics
(for a given experimental precision) except for scenarios in which new $ep$
couplings are
fortuitously suppressed ({\em e.g.}, left-right symmetric or $E_6$ models).
In addition, each
low-energy process encounters its own brand of theoretical uncertainties
which may limit the
interpretation of a given measurement in terms of new physics.

We conclude that the most thorough search for new physics using low-energy
PV would require a
program of measurements drawing upon the complementarity of different
processes. Here we
summarize the elements of this complementarity:

\begin{itemize}

\item[(a)] Both $ep$ asymmetry $\alr(^1{\hbox{H}})$ and the isotope ratios
${\cal R}_i$ are
sensitive to the {\em same} combination of new $e-q$ interactions (see
$\Delta\qwp$ of
Eqs. (\ref{eq:deltaqwp},\ref{eq:deltaone}-\ref{eq:deltatwo})).

\item[(b)] A 2-3\% determination of $\alr(^1{\hbox{H}})$ or a 0.1\%
determination of ${\cal R}_1$
would nearly double the present $\qpv({\hbox{Cs}})$ sensitivity for some
scenarios ({\em e.g.},
fermion compositeness and leptoquarks) but not others ({\em e.g.},
right-handed neutral
gauge bosons). Moreover, either of these PVES or isotope ratio measurements
would, together
with the present cesium APV result, afford a separate determination of new
$e-u$ and $e-d$
interactions.

\item[(c)] The planned measurement of the M\"oller asymmetry will provide
the best test of
lepton compositeness of any electroweak observable, exceeding the
$\Lambda_{ij}(ee)$ bounds
from LEP by nearly an order of magnitude. However, PV $ee$ scattering is
100 times less
sensitive to leptoquark and R parity-violating SUSY interactions than are
semi-leptonic
PV observables.

\item[(d)] The bounds on extra gauge bosons obtained from a one percent
determination of
$\alr(N\to\Delta)$ and $\alr(0^+,0)$ could exceed those derived from either
(a), (d), or
cesium APV.
\end{itemize}

\noindent Evidently, at least one additional measurement -- in addition to
the cesium APV
and planned M\"oller experiments -- is necessary to provide the complete
range of low-energy
information on new neutral current interactions.

From the standpoint of the {\em interpretation} of PV measurements, the
M\"oller asymmetry
provides the theoretically cleanest probe of new physics. The relevant
theoretical uncertainties
in this case are those associated with hadronic contributions to the
$Z-\gamma$ mixing tensor
\cite{Cza96} and with the (small) scattering backgrounds \cite{SLAC97}.
Neither source of
uncertainty appears to be problematic for the extraction of new physics
limits from
$\alr(ee)$.

The interpretation of semi-leptonic observables, however, requires improved
input
from atomic, nuclear, and hadron structure theory. The most challenging
theory issues lie with
the APV observables. A reduction in the cesium atomic theory uncertainty by
a factor of four would
make it comparable to the present experimental error. In this case, the
cesium new physics
sensitivity would improve by a factor of two. Whether or not such an
improvement in the atomic
theory can be achieved is an open question. In the case of APV isotope
ratios, the attainment
of the new physics sensitivity discussed above may require an experimental
determination of
$\rho_n(r)$ using PVES. At present, there exist no published estimates of
the isotope shifts in
$\rho_n$ for Yb.  Estimates for cesium and barium suggest that the
theoretical isotope shift
uncertainty may be about two times larger than desirable for future new
physics searches. In
the absence of improved nuclear theory input, measurements of the ${\cal
R}$ will provide more
information on nuclear structure than on new electroweak physics. A precise
determination of
$\rho_n$ using PVES, however, may sufficiently constrain model calculations
so as to
significantly reduce the theoretical isotope shift uncertainty.

The theoretical issues
entering the interpretation of semi-leptonic PVES appear less formidable.
The dominant corrections
to the $\qpv$ term of the asymmetry -- including both hadronic form factors
and the $\gamma\gamma$
dispersion correction -- are measurable in principle. The remaining hadron
and nuclear
structure-dependent corrections are fortuitously suppressed. Consequently,
the primary challenge
in peforming new physics searches with PVES will be experimental.

\acknowledgements

It is a pleasure to thank W.J. Marciano, D. Budker, E.N. Fortson, S.J.
Pollock, and P. Souder
for useful discussions and S.J. Puglia for assistance in preparing the
figures. This
work was supported in part under U.S. Department of Energy contract
\#DE-FG06-90ER40561
and a National Science Foundation Young Investigator Award.

\begin{figure}

\bigskip
\begin{center}
\ \epsffile{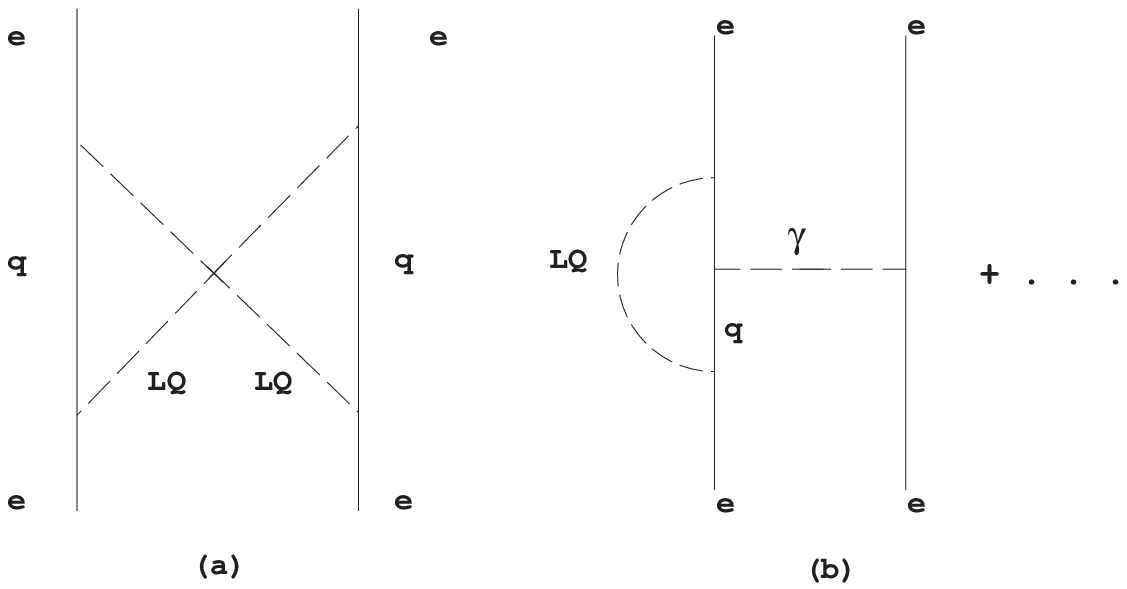}

\medskip
\caption{\label{Fig1} Leptoquark (LQ) one-loop contributions to PV M\"oller
scattering.}
\end{center}

\bigskip

\begin{center}
\ \epsffile{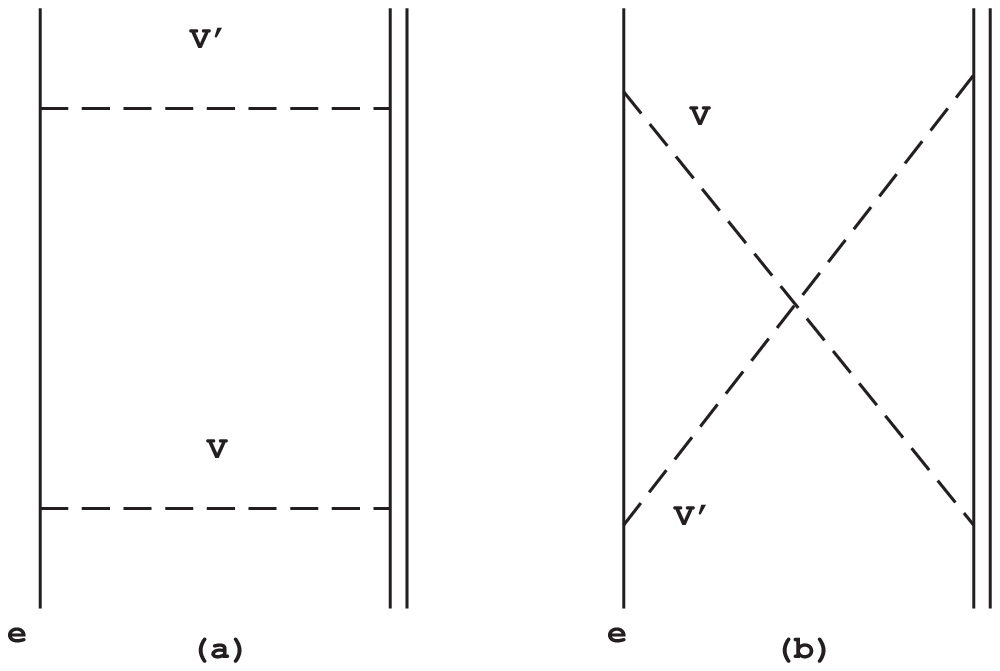}

\medskip
\caption{\label{Fig2} Two vector boson exchange dispersion corrections.
Here $V$ and $V'$
denote $\gamma$, $Z^0$, or $W^\pm$. Double vertical line denotes hadronic
target.}
\end{center}

\end{figure}

\end{document}